\documentstyle[amssymb,12pt,thmsa,sw20aip]{article}

\input tcilatex

\begin{document}

\title{Algebraic treatment of super-integrable potentials}
\author{L. Chetouani, L. Guechi \\
D\'epartement de Physique, \\
Facult\'e des Sciences, Universit\'e Mentouri,\\
Route d'Ain El Bey, Constantine, Algeria. \and T.F. Hammann \\
Laboratoire de Math\'ematiques, Physique Math\'ematique,\\
Facult\'e des Sciences et Techniques,\\
Universit\'e de Haute Alsace,\\
4, rue des Fr\`eres Lumi\`ere, F-68093 Mulhouse, France}
\date{}
\maketitle

\begin{abstract}
The so$(2,1)$ Lie algebra is applied to three classes of two- and
three-dimensional Smorodinsky-Winternitz super-integrable potentials for
which the path integral discussion has been recently presented in the
literature. We have constructed the Green's functions for two important
super-integrable potentials in $R^2.$ Among the super-integrable potentials
in $R^3$, we have considered two examples, one is maximally super-integrable
and another one minimally super-integrable. The discussion is made in
various coordinate systems. The energy spectrum and the suitably normalized
wave functions of bound and continuous states are then deduced.


\end{abstract}

\section{Introduction}

Ever since the success of the algebraic approach based on the non-compact
groups in calculating the wave functions and the transition amplitudes for
the hydrogen atom by Kleinert\cite{KLEINERT1} , a renewed interest for this
method has been emerging. Hence a large amount of physical problems for
which the Schr\"odinger equation reduces to the confluent hypergeometric
equation have been treated in the framework of this approach. In particular,
the radial harmonic oscillator, the radial Coulomb and Morse oscillator
potentials and the Natanzon confluent potentials \cite{NATANZON} , which
generalize the latter, have been discussed in a variety of SO$(2,1)$
realizations \cite{ENGLEFIELD,HUFFAKER,CIZEK,BERRONDO,WU,CORDERO,COOPER} .

The above so$(2,1)$ algebraic approach has also been applied to noncentral
potentials such as the ring-shaped potential \cite{GERRY} introduced by
Hartmann \cite{HARTMANN} to describe the molecular interaction of cyclic
polyenes and the so-called double ring-shaped oscillator \cite
{CARPIO-BERNIDO} which is the Quesne ring-shaped oscillator \cite{QUESNE}
plus an $r^{-2}\sec ^{2}\theta $ term.

The (MS) variant of the algebraic method summarized in section II has been
developed by Milshtein and Strakhovenko \cite{MILSHTEIN} to construct the
Green's function associated with the problem of Dirac electron in a static
Coulomb field. The straightforward algebraic calculation of the Green's
function for a given potential represents an undeniable advantage which
allowed this variant to be given a great deal of attention, in recent years,
following the development of path integration techniques. Hence a remarkable
set of potentials has been studied in this algebraic approach. Among them,
we can quote potentials of practical interest, such as Morse's \cite
{CHETOUANI1} , the radial harmonic oscillator and the radial Coulomb
potentials . The Hartmann ring-shaped potential \cite{VAIDYA} , the compound
Coulomb plus Aharonov-Bohm potential \cite{BOSCHI1} and two highly singular
non-isotropic potentials associated to a highly distorted spherical Coulomb
field with an additional double ring well and a highly distorted cylindrical
Coulomb field have also been studied in parabolic coordinate systems \cite
{SOUZA} . Within the framework of the $R^4$ to $R^3$ non-bijective
Kustaanheimo-Stiefel mapping, the Kaluza-Klein monopole system \cite
{CHETOUANI2} and a noncentral potential \cite{CHETOUANI3} which generalizes
the Coulomb potential and the Hartmann ring-shaped potential and also, due
to its close link with the latter, the compound Coulomb plus Aharonov-Bohm
potential have been treated by means of the same algebraic approach as
Milshtein and Strakhovenko. Cylindrical parabolic coordinates have also been
used in discussion for the so$(2,1)$ algebraic method of another type of
noncentral potentials \cite{BOSCHI2,CHETOUANI4} . All these potentials
discussed with the help of the (MS) variant appear in the general
classification of potentials in two and three dimensions possessing
dynamical invariance groups initiated about 30 years ago by Smorodinsky and
co-workers \cite{MAKAROV} , continued by Kibler and Winternitz \cite{KIBLER1}
, and revived, in recent years, by Evans \cite{EVANS1,EVANS2} . This
classification was established according to the number of degrees of
freedom, quadratic integrals of motion in the momenta and coordinate systems
in which the potential allows the separation of variables. The Hamiltonian
systems with these potentials are called super-integrable. Generally, in $n$
dimensions, a system is called ''minimally'' super-integrable if it has $%
\left( 2n-2\right) $ constants or integrals of motion ( including energy ),
and it is called ''maximally'' super-integrable if it has $\left(
2n-1\right) $ integrals of motion \cite{BONASTOS} . A list of minimally
super-integrable and maximally super-integrable potentials with the
corresponding constants of motion in the classical form and all separating
coordinate systems has been established by Evans\cite{EVANS3} . On the basis
of this classification, Grosche et al \cite{GROSCHE1} have recently
presented a detailed path integral discussion of the so-called
Smorodinsky-Winternitz super-integrable potentials in many coordinate
systems. It is to be noted that almost all the potentials contained in this
classification involve centrifugal or angular barriers which possess point
singularities. Consequently, following Kleinert \cite{KLEINERT2} the time
sliced path integral for these potentials does not exist in any coordinate
system. So, it is necessary to regularize the system in question by an
appropriate set of new coordinates in order to find a path integral
expression without collapse. This problem does not occur in the framework of
the (MS) variant of the so$(2,1)$ algebraic approach owing to its local
(differential) character. This constitutes a great advantage in studying the
problem of the Smorodinsky-Winternitz potentials in this framework.

The plan of this article is as follows. We briefly review the so$(2,1)$ Lie
algebra and its use in calculating Green's functions in section II. We study
a set of two potentials in two dimensions, algebraically obtain the Green's
function in the various separating coordinate systems and deduce the energy
spectrum as well as the corresponding normalized wave functions in section
III. Sections IV and V deal with two examples of three-dimensional
potentials. The construction of the Green's functions is made in different
coordinate systems. The energy spectrum and the normalized wave functions
are evaluated. Section VI will be a conclusion.

\section{Green's function and so$(2,1)$ Lie algebra}

Let's briefly review the main features of so$(2,1)$ Lie algebra and its use
in the calculation of the Green's functions to make this paper
self-contained. A set of three operators $\left\{ T_1,T_2,T_3\right\} $
characterized by the commutation relations \cite{GILMORE} define it:

\begin{equation}
\left[ T_1,T_2\right] =-iT_1,\quad \text{ }\left[ T_2,T_3\right]
=-iT_3,\quad \text{ }\left[ T_1,T_3\right] =-iT_2.  \label{a2.1}
\end{equation}
Because of the type of potentials we shall deal with in this paper, we have
to use the following differential realization of the operators:

\begin{equation}
T_1(x)\!=\!-\frac{\hbar ^2}{2M}\left( \!\frac{\partial ^2}{\partial x^2}\!-\!%
\frac{\mu (\mu -1)}{x^2}\!\right) ,T_2(x)\!=\!-\frac i2\left( \!x\frac
\partial {\partial x}\!+\!\frac 12\!\right) ,T_3(x)\!=\!\frac M{4\hbar
^2}x^2,  \label{a2.2}
\end{equation}
with $0<x<\infty $.

By using Schwinger's integral representation \cite{SCHWINGER} , the Green's
function associated to a potential $V(x)$ with the SO$(2,1)$ group symmetry
is given by 
\begin{equation}
G(x,x^{\prime };E)=\int_0^\infty dS\exp \left[ \frac{iS}\hbar (E+i0)\right]
K(x,x^{\prime };S),  \label{a2.3}
\end{equation}
where 
\begin{eqnarray}
K(x,x^{\prime };S) &=&\exp \left\{ -\frac{iS}\hbar \left[ -\frac{\hbar ^2}{2M%
}\nabla _x^2+V(x)\right] \right\} \delta (x-x^{\prime })  \nonumber \\
&=&\exp \left\{ -\frac{iS}\hbar \left[ T_1(x)+2\hbar ^2\omega ^2T_3(x)\right]
\right\} \delta (x-x^{\prime }).  \label{a2.4}
\end{eqnarray}
The calculation of this kernel is based upon the use of two
Baker-Campbell-Hausdorff formulas\cite{WYBOURNE} 
\begin{equation}
\exp \left\{ -\frac{iS}\hbar \left[ T_1+2\hbar ^2\omega ^2T_3\right]
\right\} =\exp (-iaT_3)\exp (-ibT_2)\exp (-icT_1),  \label{a2.5}
\end{equation}
where 
\begin{equation}
a=2\hbar \omega \tan (\omega S),\quad b=2Ln\left[ \cos (\omega S)\right]
,\quad c=\frac 1{\hbar \omega }\tan (\omega S),  \label{a2.6}
\end{equation}
and 
\begin{equation}
\exp (-i\alpha T_3)\exp (-i\beta T_2)\exp (-i\gamma T_1)=\exp (-icT_1)\exp
(\tau T_3),  \label{a2.7}
\end{equation}
with 
\begin{equation}
\alpha =\frac{i\tau }{1-i\tau c/2},\quad \beta =2Ln\left( 1-\frac{i\tau c}%
2\right) ,\quad \gamma =\frac c{1-i\tau c/2}.  \label{a2.8}
\end{equation}
Here, we also have to use the Laplace transform of the Dirac distribution 
\begin{equation}
\delta (x-x^{\prime })=\frac M{2\hbar ^2}\frac{x^\mu x^{\prime 1-\mu }}{%
2i\pi }\int_{-i\infty +\delta }^{i\infty +\delta }d\tau \exp \left[ \frac
M{4\hbar ^2}(x^2-x^{\prime 2})\tau \right] ;\quad \delta <0,  \label{a2.9}
\end{equation}
in order to obtain a manageable result as follows: 
\begin{equation}
\exp (-i\gamma T_1)x^\mu =\left[ 1-i\gamma T_1+\frac 1{2!}(-i\gamma
T_1)^2+...\right] x^\mu =x^\mu .  \label{a2.10}
\end{equation}
Using relations (\ref{a2.9}), (\ref{a2.5}) and (\ref{a2.7}), the kernel (\ref
{a2.4}) can now be written 
\begin{eqnarray}
K(x,x^{\prime };S) &=&\frac M{2\hbar ^2}x^{\prime 1-\mu }\exp (-iaT_3)\exp
(-ibT_2)x^\mu \exp \left( \frac{iMx^2}{2\hbar ^2c}\right)  \nonumber \\
&&\times \int_{-i\infty +\delta }^{i\infty +\delta }d\tau \frac{\exp \left( -%
\frac{Mx^{\prime 2}}{4\hbar ^2}\tau \right) \exp \left( \frac{Mx^2}{\hbar
^2c^2}\frac 1{\tau +2i/c}\right) }{\left( 1-\frac{i\tau c}2\right) ^{\mu
+\frac 12}},  \label{a2.11}
\end{eqnarray}
where the well-known formula 
\begin{equation}
\exp (-i\beta T_2)f(x)=\exp \left( -\frac \beta 4\right) f\left( e^{-\frac
\beta 2}x\right)  \label{a2.12}
\end{equation}
has also been used.

The integral can be calculated thanks to the residue theorem after the $\exp
\left( \frac{Mx^2}{\hbar ^2c^2}\frac 1{\tau +2i/c}\right) $ series has been
effected. Hence, we obtain 
\begin{eqnarray}
K(x,x^{\prime };S) &=&\frac{M\omega }{i\hbar \sin (\omega S)}\sqrt{%
xx^{\prime }}\exp \left[ \frac{iM\omega }{2\hbar }(x^2+x^{\prime 2})\cot
(\omega S)\right]  \nonumber \\
&&\times I_\lambda \left( \frac{M\omega xx^{\prime }}{i\hbar \sin (\omega S)}%
\right) ,  \label{a2.13}
\end{eqnarray}
where $\lambda =\mu -\frac 12$ and $I_\lambda (x)$ is the modified Bessel
function.

We can now use (\ref{a2.13}) for any coordinate in a multi-dimensional
system, provided that the $(H-E)^{-1}$ inverse resolvent operator can be
transformed into a linear combination of the above mentioned $T_i(i=1,2,3)$
operators.

\section{Two-dimensional maximally super-integrable potentials}

We shall study here a set of two important potentials belonging to a class
of two-dimensional Smorodinsky-Winternitz potentials. They are characterized
by the existence of three functionally independent integrals of motion,
which means that there is a pair of quadratic operators corresponding to
these integrals of motion which commute with the system's Hamiltonian. The
number of such integrals being superior to that of degrees of liberty, they
are thus called maximally super-integrable potentials.

\subsection{Let us study the potential}

\begin{equation}
V_1(\overrightarrow{\rho })\!=\!-\frac{\alpha _0}{\sqrt{x_1^2\!+\!x_2^2}}%
\!+\!\frac{\hbar ^2}{4M}\frac 1{\sqrt{x_1^2\!+\!x_2^2}}\left( \!\frac{%
k_1^2\!-\!\frac 14}{\sqrt{x_1^2\!+\!x_2^2}\!+\!x_1}+\frac{k_2^2\!-\!\frac 14%
}{\sqrt{x_1^2\!+\!x_2^2}\!-\!x_1}\right) ,  \label{a3.1}
\end{equation}
with positive $\alpha _0,k_1$ and $k_2$ constants . It admits the following
three functionally independent integrals of motion: 
\begin{equation}
\left\{ 
\begin{tabular}{l}
$H_1=\frac{P^2}{2M}+V_1(\overrightarrow{\rho }),\quad \quad I_1=\frac{L_3^2}{%
2M}+\frac{\hbar ^2\rho ^2}{8M}\left( \frac{k_1^2-\frac 14}{x_1^2}+\frac{%
k_2^2-\frac 14}{x_2^2}\right) $ \\ 
$I_2=\frac{\left\{ L_3,P_2\right\} }{4M}+\frac{\alpha (\eta -\xi )}{\xi \eta 
}-\frac{2(k_1^2-\frac 14)}{\xi ^2}+\frac{2(k_2^2-\frac 14)}{\eta ^2}.$%
\end{tabular}
\right.  \label{a3.2}
\end{equation}

This potential is exactly solvable in two coordinate systems, namely
parabolic and polar. For $k_1=k_2=\frac 12$, equation (\ref{a3.1}) can be
reduced to the Coulomb potential treated with a path integral approach\cite
{DURU1,INOMATA1} . The algebraic solution to this potential via the (MS)
variant is easier to establish using the parabolic coordinates or even the
Levi-Cevita variables\cite{LEVITA} defined by $x_1=u_1^2-u_2^2,\quad
x_2=2u_1u_2$ ($-\infty <u_1,u_2<+\infty $). The Green's function $G(%
\overrightarrow{\rho },\overrightarrow{\rho ^{\prime }};E)$ associated to
the potential (\ref{a3.1}), in Schwinger's integral representation, is given
by

\begin{eqnarray}
G(\overrightarrow{\rho },\overrightarrow{\rho ^{\prime }};E)\!
&=&\!\int_0^\infty \!\frac{dS}{4\rho }\!\exp \left\{ \!-\!\frac{iS}\hbar %
\left[ \!\frac 1{4\rho }\!\sum_{j=1}^2\left( \!-\frac{\hbar ^2}{2M}\left( \!%
\frac{\partial ^2}{\partial u_j^2}\!-\!\frac{k_j^2\!-\!\frac 14}{u_j^2}%
\right) \!\right) \right. \right.  \nonumber \\
&&\ \left. \left. -\frac{\alpha _0}\rho -E-i0\right] \right\}
\prod_{j=1}^2\delta (u_j-u_j^{\prime }).  \label{a3.3}
\end{eqnarray}

By applying the time transformation $S\rightarrow \tau $ defined by $\tau
=\frac S{4\rho },$ the Green's function (\ref{a3.3}) can be written 
\begin{equation}
G(\overrightarrow{\rho },\overrightarrow{\rho ^{\prime }};E)=\int_0^\infty
d\tau \exp \left[ \frac i\hbar (4\alpha _0+i0)\tau \right]
\prod_{j=1}^2K(u_j,u_j^{\prime };\tau ),  \label{a3.4}
\end{equation}
where 
\begin{equation}
K(u_j,u_j^{\prime };\tau )=\exp \left\{ -\frac{i\tau }\hbar \left[
T_1(u_j)+2\hbar ^2\omega ^2T_3(u_j)\right] \right\} \delta (u_j-u_j^{\prime
}),  \label{a3.5}
\end{equation}
with $\mu _j=\frac 12\pm k_j$ and $\omega =\sqrt{-\frac{8E}M}.$

Using the equation (\ref{a2.13}), we can write the propagators (\ref{a3.5})
as follows: 
\begin{equation}
K(u_j,u_j^{\prime };\tau )\!=\!\frac{M\omega \sqrt{u_ju_j^{\prime }}}{i\hbar
\sin (\omega \tau )}\exp \left[ \!\frac{iM\omega }{2\hbar }\left(
u_j^2\!+\!u_j^{\prime 2}\right) \!\!\cot (\omega \tau )\!\right] I_{\pm
k_j}\left( \!\frac{M\omega u_ju_j^{\prime }}{i\hbar \!\sin (\omega \tau )}%
\!\right) .  \label{a3.6}
\end{equation}

\subsubsection{Parabolic coordinates}

In the parabolic coordinates $\xi =\sqrt{2}u_1,\eta =\sqrt{2}u_2$ and after
changing $\omega \rightarrow 2\omega $ and $\tau \rightarrow \frac \tau 2$
and by taking into account (\ref{a3.6}), the Green's function (\ref{a3.4})
becomes

\begin{equation}
G(\overrightarrow{\rho },\overrightarrow{\rho ^{\prime }};E)=\int_0^\infty
d\tau \exp \left[ \frac i\hbar (2\alpha _0+i0)\tau \right] K(\xi ,\eta ,\xi
^{\prime },\eta ^{\prime };\tau ),  \label{a3.7}
\end{equation}
where 
\begin{eqnarray}
K(\xi ,\eta ,\xi ^{\prime },\eta ^{\prime };\tau ) &=&\left( \frac{M\omega }{%
i\hbar \sin (\omega \tau )}\right) ^2\sqrt{\xi \xi ^{\prime }\eta \eta
^{\prime }}  \nonumber \\
&&\times \exp \left[ \frac{iM\omega }{2\hbar }(\xi ^2+\eta ^2+\xi ^{\prime
2}+\eta ^{\prime 2})\cot (\omega \tau )\right]  \nonumber \\
&&\times I_{\pm k_1}\left( \frac{M\omega \xi \xi ^{\prime }}{i\hbar \sin
(\omega \tau )}\right) I_{\pm k_2}\left( \frac{M\omega \eta \eta ^{\prime }}{%
i\hbar \sin (\omega \tau )}\right) ,  \label{a3.8}
\end{eqnarray}
with 
\begin{equation}
\omega =\sqrt{-\frac{2E}M}.  \label{a3.9}
\end{equation}

To find the energy spectrum and the normalized wave functions of the bound
states, we make use the Hille and Hardy formula \cite{GRADSHTEIN} 
\begin{eqnarray}
\!\!\!\!\!\!\!\!\!\!\!\!\frac 1{1\!-\!z}\left[ -z\frac{x\!+\!y}{1\!-\!z}%
\right] I_\alpha \left( \!\frac{2\sqrt{xyz}}{1\!-\!z}\right) \!\!\!\!
&=&\!\!\!\!\left( xyz\right) \!^{\frac \alpha 2}\!\sum_{n=0}^\infty \!\frac{%
n!}{\Gamma (n\!+\!\alpha +\!1)}L_n^\alpha (x)L_n^\alpha (y)z^n;\mid z\mid <1,
\nonumber \\
&&  \label{a3.10}
\end{eqnarray}
where the $L_n^\alpha (x)$ are the Laguerre polynomials, and the integration
over $S$ yields the quantization condition 
\begin{equation}
n_1+n_2\pm \frac{k_1}2\pm \frac{k_2}2+p=0\text{.}  \label{a3.11}
\end{equation}
where $p=-\frac{\alpha _0}{\hbar \omega }.$ Therefore, the Green's function (%
\ref{a3.7}) can be written as: 
\begin{equation}
G(\overrightarrow{\rho },\overrightarrow{\rho ^{\prime }};E)=i\hbar
\sum_{n_1,n_2=0}^\infty \frac{\Psi _{n_1,n_2}(\xi ,\eta )\Psi
_{n_1,n_2}^{*}(\xi ^{\prime },\eta ^{\prime })}{E+i0-E_{n_1,n_2}},
\label{a3.12}
\end{equation}
with the normalized wave functions ($a=\frac{\hbar ^2}{M\alpha _0}$ is the
Bohr radius) 
\begin{eqnarray}
\Psi _{n_1,n_2}(\xi ,\eta ) &=&\left[ \frac 2{a^2N^3}\frac{n_1!n_2!}{\Gamma
(n_1\pm k_1+1)\Gamma (n_2\pm k_2+1)}\right] ^{\frac 12}\left( \frac{\xi ^2}{%
aN}\right) ^{\frac 14\pm \frac{k_1}2}  \nonumber \\
&&\left( \frac{\eta ^2}{aN}\right) ^{^{\frac 14\pm \frac{k_2}2}}\exp \left[ -%
\frac{\xi ^2+\eta ^2}{2aN}\right] L_{n_1}^{\pm k_1}\left( \frac{\xi ^2}{aN}%
\right) L_{n_2}^{\pm k_2}\left( \frac{\eta ^2}{aN}\right) ,  \nonumber \\
&&  \label{a3.13}
\end{eqnarray}
and the discrete energy spectrum given by 
\begin{equation}
E_{n_{1,}n_2}=-\frac{M\alpha _0^2}{2\hbar ^2N^2};\quad N=n_1+n_2\pm \frac{k_1%
}2\pm \frac{k_2}2+1.  \label{a3.14}
\end{equation}

To determine the energy spectrum and the wave functions of the continuous
states, let's go back to the expression (\ref{a3.8}) and use the dispersion
formula (see Ref. \cite{GRADSHTEIN} , p. 884, Eq. (7.694))

\begin{eqnarray}
&&\frac{2\pi \sqrt{xy}}{\sin \alpha }\exp \left[ -(x+y)\cot \alpha \right]
I_{2\mu }\left( \frac{2\sqrt{xy}}{\sin \alpha }\right)  \nonumber \\
&=&\int_Rdp\frac{\Gamma (\frac 12+\mu +ip)\Gamma (\frac 12+\mu -ip)}{\Gamma
^2(2\mu +1)}M_{ip,\mu }(-2ix)M_{-ip,\mu }(2iy).  \label{a3.15}
\end{eqnarray}

Given $x=\frac{M\omega }{2i\hbar }\xi ^{\prime 2},y=\frac{M\omega }{2i\hbar }%
\xi ^2$ and $\alpha =\omega \tau $ , we obtain 
\begin{eqnarray}
K(\xi ,\eta ,\xi ^{\prime },\eta ^{\prime };\tau ) &=&\frac 1{\sqrt{\xi \xi
^{\prime }\eta \eta ^{\prime }}\pi ^2\Gamma ^2(1\pm k_1)\Gamma ^2(1\pm
k_2)}\int_{-\infty }^\infty \int_{-\infty }^\infty dp_\xi dp_\eta  \nonumber
\\
&&\times \Gamma \left( \frac 12\pm \frac{k_1}2+ip_\xi \right) \Gamma \left(
\frac 12\pm \frac{k_1}2-ip_\xi \right)  \nonumber \\
&&\times \Gamma \left( \frac 12\pm \frac{k_2}2+ip_\eta \right) \Gamma \left(
\frac 12\pm \frac{k_2}2-ip_\eta \right) e^{(\pi -2\omega \tau )(p_\xi
+p_\eta )}  \nonumber \\
&&\times M_{-ip_\xi ,\pm \frac{k_1}2}\left( \frac{M\omega }\hbar \xi
^2\right) M_{ip_\xi ,\pm \frac{k_1}2}\left( \frac{M\omega }\hbar \xi
^{\prime 2}\right)  \nonumber \\
&&\times M_{-ip_\eta ,\pm \frac{k_2}2}\left( \frac{M\omega }\hbar \eta
^2\right) M_{ip_\eta ,\pm \frac{k_2}2}\left( \frac{M\omega }\hbar \eta
^{\prime 2}\right) .  \label{a3.16}
\end{eqnarray}
If we now transfer (\ref{a3.16}) into (\ref{a3.7}), we shall obtain the
poles of the continuous state Green's function by integration on the $\tau $
variable. They will be defined by 
\begin{equation}
\omega \left( p_\xi +p_\eta \right) -\frac{i\alpha _0}\hbar =0.
\label{a3.17}
\end{equation}
If we convert this into energy via (\ref{a3.9}), the values of the energy
will be 
\begin{equation}
E_p=\frac{\hbar ^2p^2}{2M}\text{ \quad with\quad }p=\frac 1{a\left( p_\xi
+p_\eta \right) }.  \label{a3.18}
\end{equation}
Let's now change the variables defined by 
\begin{equation}
p_\xi =\frac 1{2p}\left( \frac 1a+\varsigma \right) \quad \text{ and \quad }%
p_\eta =\frac 1{2p}\left( \frac 1a-\varsigma \right) .  \label{a3.19}
\end{equation}
In this case, we can write the Green's function as follows:

\begin{equation}
G^c(\xi ,\eta ,\xi ^{\prime },\eta ^{\prime };E)=i\hbar \int_0^\infty
dp\int_{-\infty }^\infty d\varsigma \frac{\Psi _{p,\varsigma }(\xi ,\eta
)\Psi _{p,\varsigma }^{*}(\xi ^{\prime },\eta ^{\prime })}{E+i0-E_p},
\label{a3.20}
\end{equation}
with the wave functions given by 
\begin{eqnarray}
\Psi _{p,\varsigma }(\xi ,\eta ) &=&\frac{\left| \Gamma \left( \frac 12\pm 
\frac{k_1}2+\frac i{2p}(\frac 1a+\varsigma )\right) \Gamma \left( \frac
12\pm \frac{k_2}2+\frac i{2p}(\frac 1a-\varsigma )\right) \right| }{2\pi
\Gamma (1\pm k_1)\Gamma (1\pm k_2)\sqrt{\xi \eta }}\frac{e^{\frac \pi {2ap}}%
}{\sqrt{p}}  \nonumber \\
&&\times M_{-\frac i{2p}(\frac 1a+\varsigma ),\pm \frac{k_1}2}(-ip\xi
^2)M_{-\frac i{2p}(\frac 1a-\varsigma ),\pm \frac{k_2}2}(-ip\eta ^2).
\label{a3.21}
\end{eqnarray}

\subsubsection{Polar coordinates}

Let's use the $(\rho ,\phi )$ polar coordinates defined by 
\begin{equation}
\xi =\sqrt{2}u_1=\sqrt{2\rho }\cos \frac \phi 2,\eta =\sqrt{2}u_2=\sqrt{%
2\rho }\sin \frac \phi 2,  \label{a3.22}
\end{equation}
and the addition theorem formula \cite{ERDELYI}: 
\begin{eqnarray}
&&\frac z2I_\nu (z\sin \alpha \sin \beta )I_\mu (z\cos \alpha \cos \beta ) 
\nonumber \\
&=&(\sin \alpha \sin \beta )^\nu (\cos \alpha \cos \beta )^\mu
\sum_{n=0}^\infty (\nu +\mu +2n+1)\frac{n!}{\Gamma (\nu +n+1)}  \nonumber \\
&&\times \frac{\Gamma (\nu +\mu +n+1)}{\Gamma (\mu +n+1)}I_{\mu +\nu
+2n+1}(z)P_n^{(\nu ,\mu )}(\cos (2\alpha ))P_n^{(\nu ,\mu )}(\cos (2\beta )).
\nonumber \\
&&  \label{a3.23}
\end{eqnarray}
This will give us the following form of the Green's function (\ref{a3.7}): 
\begin{equation}
G(\overrightarrow{\rho },\overrightarrow{\rho ^{\prime }};E)=\sum_{n=0}^%
\infty G_n(\rho ,\rho ^{\prime };E)\Phi _n^{(\pm k_2,\pm k_1)}\left( \frac
\phi 2\right) \Phi _n^{(\pm k_2,\pm k_1)}\left( \frac{\phi ^{\prime }}%
2\right) ,  \label{a3.24}
\end{equation}
where the angular wave functions are those defined in function of the 
\newline
$P_n^{(\pm k_2,\pm k_1)}(\cos \phi )$ Jacobi polynomials by 
\begin{eqnarray}
\Phi _n^{(\pm k_2,\pm k_1)}(\phi ) &=&\left[ 2(2n\pm k_1\pm k_2+1)\frac{%
n!\Gamma (n\pm k_1\pm k_2+1)}{\Gamma (n\pm k_1+1)\Gamma (n\pm k_2+1)}\right]
^{\frac 12}  \nonumber \\
&&\times (\sin \phi )^{\frac 12\pm k_2}(\cos \phi )^{\frac 12\pm
k_1}P_n^{(\pm k_2,\pm k_1)}(\cos (2\phi )).  \label{a3.25}
\end{eqnarray}
The radial Green's function $G_n(\rho ,\rho ^{\prime };E)$ included in (\ref
{a3.24}) is defined by 
\begin{eqnarray}
G_n(\rho ,\rho ^{\prime };E) &=&\frac{M\omega }{i\hbar }\int_0^\infty \frac{%
d\tau }{\sin (\omega \tau )}\exp \left[ \frac i\hbar (2\alpha _0+i0)\tau %
\right]  \nonumber \\
&&\times \exp \left[ \frac{iM\omega }{2\hbar }(\rho +\rho ^{\prime })\cot
(\omega \tau )\right] I_{2\lambda }\left( \frac{2M\omega \sqrt{\rho \rho
^{\prime }}}{i\hbar \sin (\omega \tau )}\right) ,  \label{a3.26}
\end{eqnarray}
with $\lambda =n+\frac 12(1\pm k_1\pm k_2).$

Then, thanks to formula (see Ref.\cite{GRADSHTEIN} , p. 729, Eq. (6.699.4)) 
\begin{eqnarray}
&&\int_0^\infty dq\frac{e^{-2pq}}{\sinh q}\exp \left[ -\frac 12(x+y)\coth q%
\right] I_{2\gamma }\left( \frac{\sqrt{xy}}{\sinh q}\right)  \nonumber \\
&=&\frac{\Gamma \left( p+\gamma +\frac 12\right) }{\Gamma (2\gamma +1)\sqrt{%
xy}}M_{-p,\gamma }(x)W_{-p,\gamma }(y),  \label{a3.27}
\end{eqnarray}
valid for Re$(p+\gamma +\frac 12)>0,$Re$(\gamma )>0$ and $y>x$, where $%
M_{-p,\gamma }(x)$ and $W_{-p,\gamma }(y)$ are the Whittaker functions, we
can write (\ref{a3.26}) as follows: 
\begin{equation}
G_n(\rho ,\rho ^{\prime };E)=\frac{\Gamma \left( p+\lambda +\frac 12\right) 
}{2i\omega \Gamma (2\lambda +1)\sqrt{\rho \rho ^{\prime }}}M_{-p,\lambda
}\left( \frac{2M\omega }\hbar \rho ^{\prime }\right) W_{-p,\lambda }\left( 
\frac{2M\omega }\hbar \rho \right)  \label{a3.28}
\end{equation}
with $p=-\frac{\alpha _0}{\hbar \omega },$ $\omega =\sqrt{-\frac{2E}M}$ and $%
\rho >\rho ^{\prime }$.

From (\ref{a3.28}) and (\ref{a3.24}), we can deduce that the complete
Green's function is given by 
\begin{eqnarray}
G(\overrightarrow{\rho },\overrightarrow{\rho ^{\prime }};E)
&=&\sum_{n=0}^\infty \Phi _n^{(\pm k_2,\pm k_1)}\left( \frac \phi 2\right)
\Phi _n^{(\pm k_2,\pm k_1)}\left( \frac{\phi ^{\prime }}2\right) \frac{%
\Gamma (p+\lambda +\frac 12)}{2i\omega \Gamma (2\lambda +1)\sqrt{\rho \rho
^{\prime }}}  \nonumber \\
&&\times M_{-p,\lambda }\left( \frac{2M\omega }\hbar \rho ^{\prime }\right)
W_{-p,\lambda }\left( \frac{2M\omega }\hbar \rho \right) .  \label{a3.29}
\end{eqnarray}

The normalized wave functions and the energy spectrum of the bound states
are given by expression (\ref{a3.10}) applied to $G_n(\rho ,\rho ^{\prime
};E)$ as defined in (\ref{a3.26}) provided we make the adequate change of
variables and thus obtain: 
\begin{eqnarray}
\Psi _{m,n}(\rho ,\phi ) &=&\left[ \frac{m!}{a^2(m+\lambda +\frac
12)^3\Gamma (m+2\lambda +1)}\right] ^{\frac 12}\left( \frac{2\rho }{%
a(m+\lambda +\frac 12)}\right) ^\lambda  \nonumber \\
&&\times \exp \left( -\frac \rho {a(m+\lambda +\frac 12)}\right)
L_m^{2\lambda }\left( \frac{2\rho }{a(m+\lambda +\frac 12)}\right)  \nonumber
\\
&&\times \Phi _n^{(\pm k_2,\pm k_1)}\left( \frac \phi 2\right) ,
\label{a3.30}
\end{eqnarray}
\begin{equation}
E_{m,n}=-\frac{M\alpha _0^2}{2\hbar ^2(m+\lambda +\frac 12)^2}.
\label{a3.31}
\end{equation}

In order to evaluate the contribution of the continuous spectrum to the
Green's function, let's write (\ref{a3.28}) as follows: 
\begin{eqnarray}
G_n(\rho ,\rho ^{\prime };E) &=&\frac{i\hbar }{4\pi \Gamma (2\lambda +1)%
\sqrt{\rho \rho ^{\prime }}}\oint_C\frac{dz}{E+i0-\frac{\hbar ^2z^2}{2M}}%
\Gamma (p+\lambda +\frac 12)  \nonumber \\
&&\times M_{-p,\lambda }\left( \frac{2M\omega }\hbar \rho ^{\prime }\right)
W_{-p,\lambda }\left( \frac{2M\omega }\hbar \rho \right) ,  \label{a3.32}
\end{eqnarray}
where $C$ is the closed contour, 
\begin{equation}
C:\left\{ 
\begin{array}{c}
z=k;k\in \left[ -R,R\right] \\ 
z=Re^{i\phi },\phi \in (\pi ,2\pi ).
\end{array}
\right.  \label{a3.33}
\end{equation}
At the $R\rightarrow \infty $ limit, taking the asymptotic behaviour of the
Whittaker functions (see Ref. \cite{GRADSHTEIN} , p. 1061, Eq. (9.227)) into
account, it is easy to demonstrate that the integral over the semicircle
vanishes. Which leads to: 
\begin{eqnarray}
G_n(\rho ,\rho ^{\prime };E) &=&\frac{i\hbar }{4\pi \Gamma (2\lambda +1)%
\sqrt{\rho \rho ^{\prime }}}\int_{-\infty }^\infty \frac{dk}{E+i0-\frac{%
\hbar ^2k^2}{2M}}\Gamma (p+\lambda +\frac 12)  \nonumber \\
&&\times M_{-p,\lambda }(-2ik\rho ^{\prime })W_{-p,\lambda }(-2ik\rho ).
\label{a3.34}
\end{eqnarray}
By using the following formulas (see Ref. \cite{GRADSHTEIN} , pp. 1061-2,
Eqs. (9.231.2) and (9.233.1)) 
\begin{equation}
M_{\lambda ,\mu }(z)=e^{-i\pi (\mu +\frac 12)}M_{-\lambda ,\mu }(-z),\text{
\quad with \quad }2\mu \neq -1,-2,-3,...,  \label{a3.35}
\end{equation}
and 
\begin{equation}
M_{\lambda ,\mu }(z)=\Gamma (2\mu +1)e^{i\pi \lambda }\left[ \frac{%
W_{-\lambda ,\mu }(-z)}{\Gamma (\mu -\lambda +\frac 12)}+e^{-i\pi (\mu
+\frac 12)}\frac{W_{\lambda ,\mu }(z)}{\Gamma (\mu +\lambda +\frac 12)}%
\right] ,  \label{a3.36}
\end{equation}
valid for $\arg z\in \left] -\frac{3\pi }2,\frac \pi 2\right[ ,$ and $2\mu
\neq -1,-2,-3,...,$ the expression (\ref{a3.32}) can be written 
\begin{equation}
G_n(\rho ,\rho ^{\prime };E)=\frac{i\hbar }{4\pi \Gamma ^2(2\lambda +1)\sqrt{%
\rho \rho ^{\prime }}}\int_0^\infty dE_k\frac{\Psi _{k,n}(\rho )\Psi
_{k,n}^{*}(\rho ^{\prime })}{E+i0-E_k},  \label{a3.37}
\end{equation}
where $E_k=\frac{\hbar ^2k^2}{2M}$ and the radial wave functions given by 
\begin{equation}
\Psi _{k,n}(\rho )=\left( \frac M{4\pi \hbar ^2k}\right) ^{\frac 12}\frac{%
\left| \Gamma (p+\lambda +\frac 12)\right| }{\Gamma (2\lambda +1)}\frac{e^{%
\frac{i\pi p}2}}{\sqrt{\rho }}M_{-p,\lambda }(-2ik\rho ),  \label{a3.38}
\end{equation}
with $p=-\frac i{ak}.$

\subsection{Let us now study the potential}

Let us now study the potential 
\begin{equation}
V_2(\overrightarrow{\rho })\!=\!-\!\frac{\alpha _0}{\sqrt{x_1^2\!+\!x_2^2}}%
\!+\!\frac{\beta _1\sqrt{\sqrt{x_1^2\!+\!x_2^2}\!+\!x_1}\!+\!\beta _2\sqrt{%
\sqrt{x_1^2\!+\!x_2^2}\!-\!x_1}}{\sqrt{x_1^2\!+\!x_2^2}},  \label{a3.39}
\end{equation}
with real $\beta _1$ and $\beta _2$ constants. This potential has the
following three functionally independent integrals of motion: 
\begin{equation}
\left\{ 
\begin{array}{c}
H_2=\frac{P^2}{2M}+V_2(\overrightarrow{\rho }),\quad I_1=\frac{\left\{
L_3,P_1\right\} }{4M}-\frac{\alpha (\xi -\eta )+\beta _1\eta \sqrt{\xi /2}%
-\beta _2\xi \sqrt{\eta /2}}{\xi +\eta } \\ 
I_2=\frac{\left\{ L_3,P_1\right\} }{4M}-\frac{\alpha (\xi -\eta )+(\beta
_1+\beta _2)\left( \eta \sqrt{\xi /2}-\xi \sqrt{\eta /2}\right) }{\xi +\eta }
\end{array}
\right.  \label{a3.40}
\end{equation}

For given parabolic coordinates $x_1=\frac 12(\xi ^2-\eta ^2),\quad x_2=\xi
\eta ,$ ($\xi >0$ \quad and \quad $\eta \in R$), in the Schwinger's integral
representation, the Green's function associated to this potential can be
expressed 
\begin{eqnarray}
G\left( \stackrel{\rightarrow }{\rho },\stackrel{\rightarrow }{\rho ^{\prime
}};E\right) &=&\int_0^\infty dS\exp \left[ \frac{iS}\hbar (E+i0)\right] 
\nonumber \\
&&\times \exp \left[ -\frac{iS}\hbar H_2(\xi ,\eta )\right] \frac{\delta
(\xi -\xi ^{\prime })\delta (\eta -\eta ^{\prime })}{2\rho },  \label{a3.41}
\end{eqnarray}
where 
\begin{equation}
H_2(\xi ,\eta )=-\frac{\hbar ^2}{2M}\frac 1{2\rho }\left( \frac{\partial ^2}{%
\partial \xi ^2}+\frac{\partial ^2}{\partial \eta ^2}\right) -\frac{\alpha _0%
}\rho +\frac 1\rho \left( \beta _1\xi +\beta _2\eta \right) ,  \label{a3.42}
\end{equation}
with $\rho =\frac 12(\xi ^2+\eta ^2).$

If we now perform the time transformation defined by $\sigma =\frac S{2\rho
}=\frac S{\xi ^2+\eta ^2}$ to separate the $\xi $ and $\eta $ variables and
use the mutually orthogonal parabolic coordinates $\left( \widetilde{\xi },%
\widetilde{\eta }\right) \rightarrow \left( \xi -\frac{\beta _1}E,\eta -%
\frac{\beta _2}E\right) $, we arrive at 
\begin{equation}
G\left( \stackrel{\rightarrow }{\rho },\stackrel{\rightarrow }{\rho ^{\prime
}};E\right) \!\!\!=\!\!\!\int_0^\infty \!d\sigma \exp \!\left[ \!\frac{%
i\sigma }\hbar \!\left( \!2\alpha _0\!-\!\frac{\beta _1^2\!+\!\beta _2^2}%
E\!+\!i0\right) \!\right] K(\widetilde{\xi },\widetilde{\xi }^{\prime
};\sigma )K(\widetilde{\eta },\widetilde{\eta }^{\prime };\sigma ),
\label{a3.43}
\end{equation}
where each of kernels $K(\widetilde{\xi },\widetilde{\xi }^{\prime };\sigma
) $ and $K(\widetilde{\eta },\widetilde{\eta }^{\prime };\sigma )$ can be
treated with the so$(2,1)$ Lie algebra and we have 
\begin{eqnarray}
K(x,x^{\prime };\sigma )\!\! &=&\!\!\frac 12\!\!\stackunder{\mu =0}{%
\stackrel{1}{\sum }}\exp \left\{ -\frac{i\sigma }\hbar \left[ T_1(x)+2\hbar
^2\omega ^2T_3(x)\right] \right\} \delta (x-x^{\prime })  \nonumber \\
\!\! &=&\!\!\sqrt{\frac{M\omega }{2i\pi \hbar \sin (\omega \sigma )}}%
\!\!\exp \left\{ \!\frac{iM\omega }{2\hbar }\left[ \!(x^2\!+\!x^{\prime
2})\cot (\omega \sigma )\!-\!\frac{2xx^{\prime }}{\sin (\omega \sigma )}\!%
\right] \!\right\} ,  \nonumber \\
&&  \label{a3.44}
\end{eqnarray}
with $x\equiv \left( \widetilde{\xi }\text{ or }\widetilde{\eta }\right) $
and $\omega =\sqrt{-\frac{2E}M}.$

Substituting (\ref{a3.44}) into (\ref{a3.43}), we obtain 
\begin{eqnarray}
G\left( \stackrel{\rightarrow }{\rho },\stackrel{\rightarrow }{\rho ^{\prime
}};E\right) &=&\frac{M\omega }{2i\pi \hbar }\int_0^\infty \frac{d\sigma }{%
\sin (\omega \sigma )}\exp \left[ \frac{i\sigma }\hbar \left( 2\alpha _0-%
\frac{\beta _1^2+\beta _2^2}E+i0\right) \right]  \nonumber \\
&&\!\!\!\!\!\!\!\!\!\!\!\!\!\!\!\!\times \!\exp \left\{ \!\frac{iM\omega }{%
2\hbar }\left[ \!\!\left( \!\widetilde{\xi }^2\!\!+\!\!\widetilde{\eta }%
^2\!\!+\!\!\widetilde{\xi }^{\prime 2}\!\!+\!\!\eta ^{\prime 2}\!\right)
\!\!\ \cot (\omega \sigma )\!-\!\frac{2\left( \widetilde{\xi }\widetilde{\xi 
}^{\prime }\!+\!\widetilde{\eta }\widetilde{\eta }^{\prime }\right) }{\sin
(\omega \sigma )}\!\right] \!\right\} .  \nonumber \\
&&  \label{a3.45}
\end{eqnarray}

In order to determine the energy spectrum and the normalized wave functions
of the bound states of the physical system, let's apply the Mehler formula 
\cite{MORSE} 
\begin{eqnarray}
&&\frac 1{\sqrt{1-a^2}}\exp \left\{ -\frac 1{2(1-a^2)}\left[ (x^2+x^{\prime
2})(1+a^2)-4xx^{\prime }a\right] \right\}  \nonumber \\
&=&\exp \left[ -\frac 12(x^2+x^{\prime 2})\right] \sum_{n=0}^\infty \frac
1{n!}\left( \frac a2\right) ^nH_n(x)H_n(x^{\prime }).  \label{a3.46}
\end{eqnarray}

With the help of an adequate change of variables, the poles of the Green's
function (\ref{a3.45}) will be obtained thanks to an integration over $%
\sigma $; the discrete energy spectrum is found by solving the equation 
\begin{equation}
\omega ^3-\frac{2\alpha _0}{N\hbar }\omega ^2-2\frac{\beta _1^2+\beta _2^2}{%
NM\hbar }=0,\quad \text{ or as well \quad }E_N=-\frac{M\omega _N^2}2,
\label{a3.47}
\end{equation}
with $N=n_1+n_2+1.$ $\alpha _0$ being positive, then this cubic equation has
one real root:

\begin{equation}
\omega _N=\frac{2\alpha _0}{3N\hbar }+\lambda _1+\lambda _2,  \label{a3.48}
\end{equation}
where 
\begin{equation}
\lambda _j=\!\!\left[ \!\left( \!\frac{2\alpha _0}{3N\hbar }\!\right) ^3\!+\!%
\frac{\beta _1^2\!+\!\beta _2^2}{NM\hbar }\!+\!(-1)^j\sqrt{\frac{\beta
_1^2\!+\!\beta _2^2}{NM\hbar }\!\left( \!\frac{\beta _1^2\!+\!\beta _2^2}{%
NM\hbar }\!+\!2\left( \!\frac{2\alpha _0}{3N\hbar }\!\right) ^3\right) \!}%
\right] ^{\frac 13};  \label{a3.49}
\end{equation}
with $(j=1,2).$

We may obtain the normalized wave functions of the bound states from the
residues of the integrated expression of the Green's function (\ref{a3.45}), 
\begin{eqnarray}
\Psi _{n_1,n_2}(\xi ,\eta ) &=&\frac M\hbar \left( \frac 1{2^Nn_1!n_2!N\pi }%
\stackunder{\omega \rightarrow \omega _N}{\lim }\frac{\omega ^3(\omega
^2-\omega _N^2)}{\omega ^3-\frac{2\alpha _0}{N\hbar }\omega ^2-2\frac{\beta
_1^2+\beta _2^2}{NM\hbar }}\right) ^{\frac 12}  \nonumber \\
&&\!\!\!\!\!\!\!\!\!\!\!\!\!\!\!\!\times \!\exp \left[ \!-\frac{M\omega _N}{%
2\hbar }\left( \!\widetilde{\xi }^2\!+\!\widetilde{\eta }^2\!\right) \!%
\right] H_{n_1}\!\left( \!\sqrt{\frac{M\omega _N}\hbar }\widetilde{\xi }%
\!\right) H_{n_2}\!\left( \!\sqrt{\frac{M\omega _N}\hbar }\widetilde{\eta }%
\!\right) .  \nonumber \\
&&  \label{a3.50}
\end{eqnarray}
Here, it is obvious that only states with an even total number of oscillator
quanta contribute.

To find the wave functions of the continuous states, let's go back to the
Green's function (\ref{a3.45}) and make use of the following relation 
\begin{eqnarray}
&&\ \ \ \ \frac 1{\sqrt{2\pi \sin \alpha }}\exp \left( -(x+y)\cot \alpha
\right) \exp \left( \frac{2\sqrt{xy}}{\sin \alpha }\right)  \nonumber \\
\ \ \ \ \ &=&\frac 1{\left( 2\pi \right) ^2}\int_{-\infty }^\infty dpe^{(\pi
-2\alpha )p}  \nonumber \\
&&\ \ \ \ \ \ \ \ \times \left[ \left| \Gamma \left( \frac 14+ip\right)
\right| ^2E_{-\frac 12+2ip}^{(0)}\left( e^{-i\pi /4}2\sqrt{x}\right)
E_{-\frac 12-2ip}^{(0)}\left( e^{i\pi /4}2\sqrt{y}\right) \right.  \nonumber
\\
&&\ \ \ \ \ \ \ \ +\left. \left| \Gamma \left( \frac 34+ip\right) \right|
^2E_{-\frac 12+2ip}^{(1)}\left( e^{-i\pi /4}2\sqrt{x}\right) E_{-\frac
12-2ip}^{(1)}\left( e^{i\pi /4}2\sqrt{y}\right) \right] ,  \nonumber
\label{a3.51} \\
&&  \label{a3.51}
\end{eqnarray}
which is established from the dispersion formula (\ref{a3.15}). The $E_\nu
^{(0)}(z)$ and $E_\nu ^{(1)}(z)$ are even and odd parabolic cylinder
functions with respect to the variable $z$, respectively\cite{BUCHHOLZ} .
The poles of the continuous state Green's function will be obtained by
integration on the $\tau $ variable. They will be given by 
\begin{equation}
\omega \left( p_\xi +p_\eta \right) -\frac i\hbar \left( \alpha _0-\frac{%
\beta _1^2+\beta _2^2}E\right) =0.  \label{a3.52}
\end{equation}

Then, by performing the change of variables \newline
$(p_\xi ,p_\eta )\!\!\rightarrow \!\!\left[ \!\frac 1{2p}\left( \!\frac 1{%
\widetilde{a}}\!+\!\varsigma \right) \!,\!\frac 1{2p}\left( \!\frac 1{%
\widetilde{a}}\!-\!\varsigma \right) \!\right] $ where $\widetilde{a}=\frac{%
\hbar ^2}{M\left( \alpha _0-M\left( \beta _1^2+\beta _2^2\right) /\hbar
^2p^2\right) }$, it is possible to write the contribution of the continuous
part to the Green's function as 
\begin{equation}
G^c(\xi ,\eta ,\xi ^{\prime },\eta ^{\prime };E)=i\hbar \int_{-\infty
}^\infty dp\int_{-\infty }^\infty d\varsigma \frac{\Psi _{p,\varsigma
}^{*}(\xi ^{\prime },\eta ^{\prime })\Psi _{p,\varsigma }(\xi ,\eta )}{%
E+i0-E_p},  \label{a3.53}
\end{equation}
where $E_p=-\frac{M\omega ^2}2=\frac{\hbar ^2p^2}{2M}$, and the continuous
functions $\Psi _{p,\varsigma }(\xi ,\eta )$ have the form 
\begin{eqnarray}
\!\!\!\!\!\!\Psi _{p,\varsigma }(\xi ,\eta )\! &=&\!\frac{e^{\pi /2%
\widetilde{a}p}}{4\pi \sqrt{2}}\left( \! 
\begin{array}{c}
\Gamma \left( \!\frac 14\!+\!\frac i{2p}\left( \!\frac 1{\widetilde{a}%
}\!+\!\varsigma \!\right) \!\right) E_{-\frac 12+\frac ip\left( \frac 1{%
\widetilde{a}}+\varsigma \right) }^{(0)}\!\left( \!e^{-i\pi /4}\!\sqrt{2p}%
\widetilde{\xi }\!\right) \\ 
\Gamma \left( \!\frac 34\!+\!\frac i{2p}\left( \!\frac 1{\widetilde{a}%
}\!+\!\varsigma \!\right) \!\right) E_{-\frac 12+\frac ip\left( \frac 1{%
\widetilde{a}}+\varsigma \right) }^{(1)}\!\left( \!e^{-i\pi /4}\!\sqrt{2p}%
\widetilde{\xi }\!\right)
\end{array}
\!\right)  \nonumber \\
&&\times \left( \! 
\begin{array}{c}
\Gamma \left( \!\frac 14\!+\!\frac i{2p}\left( \!\frac 1{\widetilde{a}%
}\!-\!\varsigma \right) \!\right) E_{-\frac 12+\frac ip\left( \frac 1{%
\widetilde{a}}-\varsigma \right) }^{(0)}\left( \!e^{-i\pi /4}\!\sqrt{2p}%
\widetilde{\eta }\!\right) \\ 
\Gamma \left( \!\frac 34\!+\!\frac i{2p}\left( \!\frac 1{\widetilde{a}%
}\!-\!\varsigma \!\right) \!\right) E_{-\frac 12+\frac ip\left( \frac 1{%
\widetilde{a}}-\varsigma \right) }^{(1)}\left( \!e^{-i\pi /4}\!\sqrt{2p}%
\widetilde{\eta }\!\right)
\end{array}
\!\right) .  \label{a3.54}
\end{eqnarray}

\section{Three-dimensional maximally super-integrable potentials}

In three-dimensional Euclidean space, Smorodinsky and co-workers have found
a set of five potentials which have five functionally independent integrals
of motion. These three-dimensional potentials are called maximally
super-integrable potentials. At least, each potential of the so-called class
of Smorodinsky-Winternitz potentials can be treated in two coordinate
systems through the so$(2,1)$ Lie algebraic approach. Here, we shall
restrict ourselves to study the potential 
\begin{equation}
V_3(\overrightarrow{r})=-\frac{\alpha _0}r+\frac{\hbar ^2}{2M}\left( \frac{%
k_1^2-\frac 14}{x_1^2}+\frac{k_2^2-\frac 14}{x_2^2}\right) .  \label{a4.1}
\end{equation}
The integrals of motion are 
\begin{equation}
\left\{ 
\begin{tabular}{l}
$H_3=\frac{p^2}{2M}+V_3(\overrightarrow{r}),\quad I_1=\frac{L_3^2}{2M}+\frac{%
\hbar ^2}{2M}\left( \frac{k_1^2-\frac 14}{\cos ^2\phi }+\frac{k_2^2-\frac 14%
}{\sin ^2\phi }\right) ,$ \\ 
$I_2=\frac{L_2^2}{2M}+\frac{\hbar ^2}{2M}\frac{k_1^2-\frac 14}{\tan ^2\theta
\cos ^2\phi },\quad I_3=\frac{\overrightarrow{L}^2}{2M}+\frac{\hbar ^2}{%
2M\sin ^2\theta }\left( \frac{k_1^2-\frac 14}{\cos ^2\phi }+\frac{%
k_2^2-\frac 14}{\sin ^2\phi }\right) ,$ \\ 
$I_4=\frac 1{4M}\left( I_{x_1x_2}+I_{x_2x_1}\right) -(\xi -\eta )\left[
\frac \alpha {\xi +\eta }-\frac{\hbar ^2}{2M\xi \eta }\left( \frac{%
k_1^2-\frac 14}{\cos ^2\phi }+\frac{k_2^2-\frac 14}{\sin ^2\phi }\right) 
\right] ,$%
\end{tabular}
\right.  \label{a4.2}
\end{equation}
where $I_{ij}=\left\{ L_i,P_j\right\} =L_iP_j+P_jL_i$ with $(i,j)\equiv
(x_1,x_2,x_3).$

This potential is a generalization of the Coulomb potential analyzed by
various authors in the path integral \cite
{DURU2,HO,PAK,INOMATA2,STEINER,CHETOUANI5,CHETOUANI6,KLEINERT3,CASTRIGIANO,STORCHAK,GROSCHE2,CARPIO-BERNIDO2}
and algebraic approach \cite{KIBLER2,KIBLER3} contexts. As it features
singularities for $x_1=0$ and $x_2=0$, all we need is to discuss it in the $%
0<x_1,x_2<\infty $ and $x_3\in R$ area. It is possible to evaluate the
Green's function 
\begin{equation}
G(\overrightarrow{r},\overrightarrow{r^{\prime }};E)=\int_0^\infty dS\exp %
\left[ -\frac{iS}\hbar (H_3-E-i0)\right] \delta (\overrightarrow{r}-%
\overrightarrow{r^{\prime }})  \label{a4.3}
\end{equation}
in parabolic and spherical coordinates.

\subsubsection{Parabolic coordinates}

For given parabolic coordinates $x_1=\xi \eta \cos \phi ,$ $x_2=\xi \eta
\sin \phi ,$\newline
$x_3=\frac 12(\xi ^2-\eta ^2),$ $\xi ,\eta >0$ and $0\leq \phi <2\pi ,$ the
Green's function (\ref{a4.3}) can be written 
\begin{equation}
G(\overrightarrow{r},\overrightarrow{r^{\prime }};E)\!\!=\!\!\int_0^\infty
\!dS\exp \!\left[ \!-\frac{iS}\hbar (\widetilde{H}_3\!-\!E\!-\!i0)\!\right]
\!\frac{\delta (\xi \!-\!\xi ^{\prime })\delta (\eta \!-\!\eta ^{\prime
})\delta (\phi \!-\!\phi ^{\prime })}{\xi \eta (\xi ^2+\eta ^2)},
\label{a4.4}
\end{equation}
with 
\begin{eqnarray}
\widetilde{H}_3 &=&-\frac{\hbar ^2}{2M}\left[ \frac 1{\xi ^2+\eta ^2}\left( 
\frac{\partial ^2}{\partial \xi ^2}+\frac 1\xi \frac \partial {\partial \xi
}+\frac{\partial ^2}{\partial \eta ^2}+\frac 1\eta \frac \partial {\partial
\eta }\right) +\frac 1{\xi ^2\eta ^2}\frac{\partial ^2}{\partial \phi ^2}%
\right]  \nonumber  \label{a4.5} \\
&&\ -\frac{2\alpha _0}{\xi ^2+\eta ^2}+\frac{\hbar ^2}{2M}\frac 1{\xi ^2\eta
^2}\left( \frac{k_1^2-\frac 14}{\cos ^2\phi }+\frac{k_2^2-\frac 14}{\sin
^2\phi }\right) .  \label{a4.5}
\end{eqnarray}
Separating the angular part of the expression (\ref{a4.4}) by a time
transformation defined by $\tau =\frac S{\xi ^2\eta ^2}$, we can deduce that 
\begin{equation}
G(\overrightarrow{r},\overrightarrow{r^{\prime }},E)=\int_0^\infty d\tau
K(\xi ,\eta ,\xi ^{\prime },\eta ^{\prime };\tau )K(\phi ,\phi ^{\prime
};\tau ),  \label{a4.6}
\end{equation}
where 
\begin{eqnarray}
K(\xi ,\eta ,\xi ^{\prime },\eta ^{\prime };\tau ) &=&\exp \left\{ -\frac{%
i\tau }\hbar \left[ H_3(\xi ,\eta )-E\xi ^2\eta ^2-i0\right] \right\} 
\nonumber \\
&&\times \frac{\xi \eta }{\xi ^2+\eta ^2}\delta (\xi -\xi ^{\prime })\delta
(\eta -\eta ^{\prime }),  \label{a4.7}
\end{eqnarray}
and 
\begin{eqnarray}
K(\phi ,\phi ^{\prime };\tau )\! &=&\!\exp \left\{ \!-\frac{i\tau }\hbar 
\frac{\hbar ^2}{2M}\left( \!-\frac{\partial ^2}{\partial \phi ^2}\!+\!\frac{%
k_1^2\!-\!\frac 14}{\cos ^2\phi }\!+\!\frac{k_2^2\!-\!\frac 14}{\sin ^2\phi }%
\!\right) \!\right\} \!\delta (\phi \!-\!\phi ^{\prime })  \nonumber \\
&=&\sum_{n=0}^\infty \Phi _n^{(\pm k_2,\pm k_1)}(\phi )\Phi _n^{(\pm k_2,\pm
k_1)}(\phi ^{\prime })\exp \left( -\frac i\hbar \frac{\hbar ^2\lambda _1^2}{%
2M}\tau \right) ,  \label{a4.8}
\end{eqnarray}
with $\lambda _1=2n\pm k_1\pm k_2+1.$

It is to be noted that the explicit construction of the kernel $K(\phi ,\phi
^{\prime };\tau )$ thanks the (MS) variant of the algebraic technique is
being investigated and will be the subject of our forthcoming publication.

Introducing (\ref{a4.8}) into (\ref{a4.6}) and applying the $\tau
\rightarrow S$ inverse time transformation will allow us to write (\ref{a4.6}%
) as follows: 
\begin{eqnarray}
G(\overrightarrow{r},\overrightarrow{r^{\prime }};E)\!
&=&\!\sum_{n=0}^\infty \!\Phi _n^{(\pm k_2,\pm k_1)}(\phi )\Phi _n^{(\pm
k_2,\pm k_1)}(\phi ^{\prime })\!\int_0^\infty \!dS\exp \!\left[ \!\frac{iS}%
\hbar (E\!-\!i0)\right]  \nonumber \\
&&\times \exp \left\{ -\frac{iS}\hbar \left[ \widetilde{H}_3(\xi ,\eta )%
\right] \right\} \frac{\delta (\xi -\xi ^{\prime })\delta (\eta -\eta
^{\prime })}{\xi \eta (\xi ^2+\eta ^2)},  \label{a4.9}
\end{eqnarray}
where 
\begin{equation}
\widetilde{H}_3(\xi ,\eta )\!=\!-\frac{\hbar ^2}{2M}\frac 1{\xi ^2\!+\!\eta
^2}\left( \!\frac{\partial ^2}{\partial \xi ^2}\!+\!\frac 1\xi \frac
\partial {\partial \xi }\!+\!\frac{\partial ^2}{\partial \eta ^2}\!+\!\frac
1\eta \frac \partial {\partial \eta }\!\right) \!-\!\frac{2\alpha _0}{\xi
^2\!+\!\eta ^2}\!+\!\frac{\hbar ^2\lambda _1^2}{2M\xi ^2\eta ^2}.
\label{a4.10}
\end{equation}

If we now eliminate the $\frac 1\xi \frac \partial {\partial \xi }$ and $%
\frac 1\eta \frac \partial {\partial \eta }$ operators by applying $\frac{%
\partial ^2}{\partial \xi ^2}+\frac 1\xi \frac \partial {\partial \xi }$ and 
$\frac{\partial ^2}{\partial \eta ^2}+\frac 1\eta \frac \partial {\partial
\eta }$ on $\frac{\delta (\xi -\xi ^{\prime })}{\sqrt{\xi \xi ^{\prime }}}$
and $\frac{\delta (\eta -\eta ^{\prime })}{\sqrt{\eta \eta ^{\prime }}}$
respectively, we can then proceed with a new $S\rightarrow \sigma $ time
transformation defined by $\sigma =\frac S{\xi ^2+\eta ^2},$ which will
allow us to write 
\begin{eqnarray}
G(\overrightarrow{r},\overrightarrow{r^{\prime }};E)\!
&=&\!\sum_{n=0}^\infty \!\Phi _n^{(\pm k_2,\pm k_1)}(\phi )\Phi _n^{(\pm
k_2,\pm k_1)}(\phi ^{\prime })\!\int_0^\infty \!d\sigma \exp \!\left[
\!\frac i\hbar (2\alpha _0\!+\!i0)\sigma \!\right]  \nonumber \\
&&\times K_n(\xi ,\xi ^{\prime };\sigma )K_n(\eta ,\eta ^{\prime };\sigma ),
\label{a4.11}
\end{eqnarray}
with 
\begin{equation}
K_n(u,u^{\prime };\sigma )\!=\!\frac 1{\sqrt{uu^{\prime }}}\exp \left\{ \!-%
\frac{i\sigma }\hbar \left[ \!-\frac{\hbar ^2}{2M}\left( \!\frac{\partial ^2%
}{\partial u^2}\!-\!\frac{\lambda _1^2\!-\!\frac 14}{u^2}\!\right)
\!-\!Eu^2\!\right] \!\right\} \delta (u\!-\!u^{\prime }),  \label{a4.12}
\end{equation}
where $u\equiv (\xi $ or $\eta ).$

So, it is possible to give the kernel (\ref{a4.12}) in function of the so$%
(2,1)$ Lie algebra operators. Indeed, by following the equations (\ref{a2.2}%
) and (\ref{a2.13}), we obtain 
\begin{eqnarray}
K_n(u,u^{\prime };\sigma ) &=&\frac 1{\sqrt{uu^{\prime }}}\exp \left\{ -%
\frac{i\sigma }\hbar \left[ T_1(u)+2\hbar ^2\omega ^2T_3(u)\right] \right\}
\delta (u-u^{\prime })  \nonumber \\
\!\!\! &=&\!\frac{M\omega }{i\hbar \sin (\omega \sigma )}\!\exp \!\left[ \!%
\frac{iM\omega }{2\hbar }(u^2\!+\!u^{\prime 2})\!\cot (\omega \sigma )\!%
\right] I_{\lambda _1}\!\left( \!\frac{M\omega uu^{\prime }}{i\hbar \sin
(\omega \sigma )}\!\right) ,  \nonumber \\
&&  \label{a4.13}
\end{eqnarray}
with $\omega =\sqrt{-\frac{2E}M}.$ The Green's function (\ref{a4.11}) can
now be written 
\begin{equation}
G(\overrightarrow{r},\overrightarrow{r^{\prime }};E)=\sum_{n=0}^\infty \Phi
_n^{(\pm k_2,\pm k_1)}(\phi )\Phi _n^{(\pm k_2,\pm k_1)}(\phi ^{\prime
})G_n(\xi ,\eta ,\xi ^{\prime },\eta ^{\prime };E),  \label{a4.14}
\end{equation}
with 
\begin{eqnarray}
G_n(\xi ,\eta ,\xi ^{\prime },\eta ^{\prime };E) &=&\left( \frac{M\omega }{%
i\hbar }\right) ^2\int_0^\infty \frac{d\sigma }{\sin ^2(\omega \sigma )}\exp %
\left[ \frac i\hbar (2\alpha _0+i0)\sigma \right]  \nonumber \\
&&\times \exp \left[ \frac{iM\omega }{2\hbar }(\xi ^2+\eta ^2+\xi ^{\prime
2}+\eta ^{\prime 2})\cot (\omega \sigma )\right]  \nonumber \\
&&\times I_{\lambda _1}\left( \frac{M\omega \xi \xi ^{\prime }}{i\hbar \sin
(\omega \sigma )}\right) I_{\lambda _1}\left( \frac{M\omega \eta \eta
^{\prime }}{i\hbar \sin (\omega \sigma )}\right) .  \label{a4.15}
\end{eqnarray}

Thanks to the Hille and Hardy formula (\ref{a3.10}) for the discrete part
and thanks to the scattering relation (\ref{a3.15}) for the continuous part,
the Green's function (\ref{a4.14}) can be developed into partial waves as
follows: 
\begin{eqnarray}
G(\overrightarrow{r},\overrightarrow{r^{\prime }};E) &=&i\hbar
\sum_{n=0}^\infty \left\{ \sum_{n_1,n_2=0}^\infty \frac{\Psi _{n_1n_2,n}(\xi
,\eta ,\phi )\Psi _{n_1,n_2,n}^{*}(\xi ^{\prime },\eta ^{\prime },\phi
^{\prime })}{E+i0-E_N}\right.  \nonumber \\
&&\left. +\int_0^\infty dp\int_{-\infty }^\infty d\kappa \frac{\Psi
_{p,\kappa ,n}(\xi ,\eta ,\phi )\Psi _{p,\kappa ,n}^{*}(\xi ^{\prime },\eta
^{\prime },\phi ^{\prime })}{E+i0-E_p}\right\} .  \label{a4.16}
\end{eqnarray}
Hence, for bound states-the normalized wave functions and energy spectrum
will be 
\begin{eqnarray}
\Psi _{n_1,n_2,n}(\xi ,\eta ,\phi )\! &=&\!\left[ \!\frac 2{a^3N^4}\frac{%
n_1!n_2!}{\!\Gamma (n_1\!+\lambda _1\!+\!1)\Gamma (n_2\!+\!\lambda _2\!+\!1)}%
\right] ^{\frac 12}\!\left( \!\frac{\xi \eta }{aN}\!\right) ^{\lambda _1} 
\nonumber \\
&&\times \exp \left( -\frac{\xi ^2+\eta ^2}{2aN}\right) L_{n_1}^{\lambda
_1}\left( \frac{\xi ^2}{aN}\right) L_{n_2}^{\lambda _1}\left( \frac{\eta ^2}{%
aN}\right) \Phi _n^{(\pm k_2,\pm k_1)}(\phi ),  \nonumber \\
&&  \label{a4.17}
\end{eqnarray}
\begin{equation}
E_N=-\frac{M\alpha _0^2}{2\hbar ^2N^2},\quad N=n_1+n_2+\lambda _1+1,
\label{a4.18}
\end{equation}
and, for the continuous states, the normalized wave functions and the energy
spectrum will be 
\begin{eqnarray}
\Psi _{p,\kappa ,n}(\xi ,\eta ,\phi )\! &=&\!\frac{\left| \Gamma \left(
\!\frac 12\!+\!\frac{\lambda _1}2\!+\!\frac i{2p}(\frac 1a\!+\!\kappa
)\right) \Gamma \left( \!\frac 12\!+\!\frac{\lambda _1}2\!+\!\frac
i{2p}(\frac 1a\!-\!\kappa )\right) \right| }{2\pi \Gamma ^2(\lambda _1+1)\xi
\eta }  \nonumber \\
&&\times \frac{e^{\frac \pi {2ap}}}{\sqrt{p}}M_{-\frac i{2p}(\frac 1a+\kappa
),\frac{\lambda _1}2}(-ip\xi ^2)M_{-\frac i{2p}(\frac 1a-\kappa ),\frac{%
\lambda _1}2}(-ip\eta ^2)  \nonumber \\
&&\Phi _n^{(\pm k_2,\pm k_1)}(\phi ),  \label{a4.19}
\end{eqnarray}
\begin{equation}
E_p=\frac{\hbar ^2p^2}{2M}.  \label{a4.20}
\end{equation}

\subsubsection{Spherical coordinates}

To study the problem in this coordinate system, we shall use the expression
of the partial Green's function (\ref{a4.15}) and use the following change
of variables 
\begin{equation}
(\xi ,\eta )\rightarrow \left( \sqrt{2r}\cos \frac \theta 2,\sqrt{2r}\sin
\frac \theta 2\right) .  \label{a4.21}
\end{equation}
The partial Green's function (\ref{a4.15}) then can be written 
\begin{eqnarray}
G_n(r,\theta ,r^{\prime },\theta ^{\prime };E)\!\! &=&\!\!\sum_{m=0}^\infty
\!(m\!+\!\lambda _1\!+\!\frac 12)\frac{\Gamma (m\!+\!2\lambda _1\!+\!1)}{m!}%
P_{m+\lambda _1}^{-\lambda _1}(\cos \theta )  \nonumber \\
&&\!\!\!\times P_{m+\lambda _1}^{-\lambda _1}(\cos \theta ^{\prime })\frac{%
2M\omega }{i\hbar \sqrt{rr^{\prime }}}\!\int_0^\infty \!\frac{d\sigma }{\sin
(\omega \sigma )}\!\exp \!\left[ \!\frac i\hbar (2\alpha \!+\!i0)\sigma %
\right]  \nonumber \\
&&\times \exp \left[ \frac{iM\omega }\hbar (r+r^{\prime })\cot (\omega
\sigma )\right] I_{2m+2\lambda _1+1}\left( \frac{2M\omega \sqrt{rr^{\prime }}%
}{i\hbar \sin (\omega \sigma )}\right) .  \nonumber \\
&&  \label{a4.22}
\end{eqnarray}
We have used here the addition theorem (\ref{a3.23}) , the connection
between hypergeometric functions and the Jacobi polynomials (see Ref. \cite
{GRADSHTEIN} , p. 1036, Eq. (8.962)) 
\begin{equation}
_2F_1(l+\alpha +\beta +1,-l;1+\alpha ;\frac{1-t}2)=\frac{l!\Gamma (\alpha +1)%
}{\Gamma (l+\alpha +1)}P_l^{(\alpha ,\beta )}(t),  \label{a4.23}
\end{equation}
as well as the relation between the hypergeometric functions and the
Legendre polynomials (see Ref. \cite{GRADSHTEIN} , p. 1009, Eq. (8.771)),
and eventually the link between $P_\nu ^{-m}(x)$ and $P_\nu ^m(x)$ (see Ref. 
\cite{GRADSHTEIN} , p. 1008, Eq. (8.752.2)).

In order to perform the integration on the $\sigma $ time variable, we shall
use the formula (\ref{a3.27}). Consequently, the final expression of the
Green's function in spherical coordinates will be 
\begin{eqnarray}
G(\overrightarrow{r},\overrightarrow{r^{\prime }};E)\!\!
&=&\!\!\sum_{n=0}^\infty \sum_{m=0}^\infty (m\!+\!\lambda _1\!+\!\frac 12)%
\frac{\Gamma (m\!+\!\lambda _1\!+\!1)}{m!}\frac{\Gamma (p\!+\!m\!+\!\lambda
_1\!+\!1)}{\Gamma (2m\!+\!2\lambda _1\!+\!2)}  \nonumber \\
&&\!\!\!\!\times \frac 1{i\omega rr^{\prime }}M_{-p,m+\lambda _1+\frac
12}\left( \!\frac{2M\omega }\hbar r^{\prime }\!\right) \!W_{-p,m+\lambda
_1+\frac 12}\left( \!\frac{2M\omega }\hbar r\!\right)  \nonumber \\
&&\times P_{m+\lambda _1}^{-\lambda _1}(\cos \theta )P_{m+\lambda
_1}^{-\lambda _1}(\cos \theta ^{\prime })\Phi _n^{(\pm k_2,\pm k_1)}(\phi
)\Phi _n^{(\pm k_2,\pm k_1)}(\phi ^{\prime }),  \nonumber \\
&&  \label{a4.24}
\end{eqnarray}
where $r>r^{\prime }$ and $p=-\frac{\alpha _0}{\hbar \omega }.$

The Hille and Hardy formula (\ref{a3.10}) and an analytic proceeding
consisting in using the Sommerfeld-Watson transformation \cite{THIRRING}
will help us to write the Green's function (\ref{a4.24}) in the form of a
partial wave development consisting of two contributions of a discrete and a
continuous part: 
\begin{eqnarray}
G(\overrightarrow{r},\overrightarrow{r^{\prime }};E) &=&i\hbar
\sum_{n,m=0}^\infty \left\{ \sum_{l=0}^\infty \frac{\Psi _{l,m,n}(r,\theta
,\phi )\Psi _{l,m,n}^{*}(r^{\prime },\theta ^{\prime },\phi ^{\prime })}{%
E+i0-E_N}\right.  \nonumber \\
&&\ \left. +\int_0^\infty dk\frac{\Psi _{k,m,n}(r,\theta ,\phi )\Psi
_{k,m,n}^{*}(r^{\prime },\theta ^{\prime },\phi ^{\prime })}{E+i0-\frac{%
\hbar ^2k^2}{2M}}\right\} ,  \label{a4.25}
\end{eqnarray}
with-for bound states-the poles located around the values of $E$ and the
normalized wave functions respectively given by 
\begin{equation}
E_N=-\frac{M\alpha _0^2}{2\hbar ^2N^2};\quad N=l+m+\lambda _1+1,
\label{a4.26}
\end{equation}

\begin{eqnarray}
\Psi _{l,m,n}(r,\theta ,\phi )\!\! &=&\!\!\left[ \!\frac{(m\!+\!\lambda
_1\!+\!\frac 12)l!\Gamma (m\!+\!\lambda _1\!+\!1)}{a^3(l\!+\!\lambda
_2\!+\!\frac 12)^4m!\Gamma (l\!+\!2\lambda _2\!+\!1)}\!\right] ^{\frac
12}\!\left( \!\frac{2r}{a(l\!+\!\lambda _2\!+\!\frac 12)}\!\right) ^{\lambda
_2-\frac 12}  \nonumber \\
&&\times \exp \left( -\frac r{a(l+\lambda _2+\frac 12)}\right) L_l^{2\lambda
_2}\left( \frac{2r}{a(l+\lambda _2+\frac 12)}\right)  \nonumber \\
&&\times P_{m+\lambda _1}^{-\lambda _1}(\cos \theta )\Phi _n^{(\pm k_2,\pm
k_1)}(\phi ).  \label{a4.27}
\end{eqnarray}
For continuous states, the normalized wave functions and energy spectrum are
respectively given by 
\begin{eqnarray}
\Psi _{k,m,n}(r,\theta ,\phi ) &=&\frac 1{\sqrt{2\pi }}\left[ (m+\lambda
_1+\frac 12)\frac{\Gamma (m+\lambda _1+1)}{m!}\right] ^{\frac 12}\frac{%
\left| \Gamma \left( \frac 12+\lambda _2-\frac i{ak}\right) \right| }{\Gamma
(2\lambda _2+1)}  \nonumber \\
&&\times \frac{e^{\frac \pi {2ak}}}rM_{\frac i{ak},\lambda
_2}(-2ikr)P_{m+\lambda _1}^{-\lambda _1}(\cos \theta )\Phi _n^{(\pm k_2,\pm
k_1)}(\phi ),  \label{a4.28}
\end{eqnarray}

\section{Three-dimensional minimally super-integrable potentials}

There are nine three-dimensional potentials which belong to the class of
minimally super-integrable Smorodinsky-Winternitz potentials, that is to say
three-dimensional potentials characterized by the existence of four
functionally independent integrals of motion. Among them are seven
potentials which have SO$(2,1)$ as a dynamical group and thus their exact
solution can be given via the (M-S) variant of the algebraic approach in
different coordinate systems. As an example, we shall discuss the potential 
\begin{equation}
V_4(\overrightarrow{r})=-\frac{\alpha _0}r+\frac{\hbar ^2}{2M(x_1^2+x_2^2)}%
\left( \frac{k_1^2x_3}r+F\left( \frac{x_2}{x_1}\right) \right) ,
\label{a5.1}
\end{equation}
with $k_1$ a positive constant. The corresponding observables have the form 
\begin{equation}
\left\{ 
\begin{array}{c}
H_4=\frac{\overrightarrow{P}^2}{2M}+V_4(\overrightarrow{r}),\text{ }I_1=%
\frac{L_z^2}{2M}+F(\tan \phi ),\text{ } \\ 
I_2=\frac{P_z^2}{2M}+\frac{\hbar ^2}{2M}\frac{k_1^2\cos \theta +F(\tan \phi )%
}{\sin ^2\theta }, \\ 
I_3=\frac 1{4M}(I_{x_1x_2}+I_{x_2x_1})-\alpha _0\frac{\xi -\eta }{\xi +\eta }%
+\frac{\hbar ^2}{2M}\left( \frac 1\eta -\frac 1\xi \right) (k_1^2+F(\tan
\phi )),
\end{array}
\right.  \label{a5.2}
\end{equation}
where $I_{ij}=\left\{ L_i,P_j\right\} =L_iP_j+P_jL_i$ with $(i,j)\equiv
(x_1,x_2,x_3)$

The Green's function for this potential can be explicitly evaluated in the
parabolic and spherical coordinate systems. For $F\left( \frac yx\right)
=\gamma ^2$ and $\gamma $ a real constant, this potential reduces to the
ring-shaped potential proposed by Hartmann as a model for the ring-shaped
molecules. It has been analyzed by many authors in the framework of path
integrals \cite
{CARPIO-BERNIDO3,CARPIO-BERNIDO4,SOKMEN1,CHETOUANI7,CARPIO-BERNIDO5} and
through the algebraic approach \cite
{KIBLER4,KIBLER5,CALOGERO,KIBLER6,GUHA,GRANOVSKY,ZHEDANOV,KIBLER7} .We can
also notice the close link of the latter with the Coulomb potential plus the
barrier created by the solenoid of Aharonov-Bohm \cite{AHARONOV} treated
with the path integrals\cite{CHETOUANI8,SOKMEN2,LEVAN} and via the algebraic
technique\cite{KIBLER8} . To give the solution for the potential (\ref{a5.1}%
) via the (MS) variant of the so$(2,1)$ algebraic approach, we will use the
Kustaanheimo-Stiefel transformation \cite{KUSTAANHEIMO} $\left\{
x_1,x_2,x_3\right\} \rightarrow \left\{ u_{j,}j\in \left[ 1,4\right]
\right\} $ corresponding to the surjection $R^4\rightarrow R^3$, which can
be defined as: 
\begin{equation}
\left( 
\begin{array}{c}
x_1 \\ 
x_2 \\ 
x_3
\end{array}
\right) =\left( A\right) \left( 
\begin{array}{c}
u_1 \\ 
u_2 \\ 
u_3 \\ 
u_4
\end{array}
\right) ,\quad \left( A\right) =\!\left( \! 
\begin{array}{cccc}
u_3 & u_4 & u_1 & u_2 \\ 
-u_4 & u_3 & u_2 & -u_1 \\ 
-u_1 & -u_2 & u_3 & u_4 \\ 
-u_2 & u_1 & -u_4 & u_3
\end{array}
\!\right)  \label{a5.3}
\end{equation}
with the constraint 
\begin{equation}
dx_4=2(-u_2du_1+u_1du_2-u_4du_3+u_3du_4)=0,  \label{a5.4}
\end{equation}
allowing to define a fourth coordinate 
\begin{equation}
x_4=2\int^s(-u_2\stackrel{.}{u}_1+u_1\stackrel{.}{u}_2-u_4\stackrel{.}{u}%
_3+u_3\stackrel{.}{u}_4)ds.  \label{a5.5}
\end{equation}

Moreover, we can show that 
\begin{equation}
\left( 
\begin{array}{c}
\frac \partial {\partial x_1} \\ 
\frac \partial {\partial x_2} \\ 
\frac \partial {\partial x_3} \\ 
\frac 1{2r}\frac \partial {\partial x_4}
\end{array}
\right) =\frac 1{2r}\left( A\right) \left( 
\begin{array}{c}
\frac \partial {\partial u_1} \\ 
\frac \partial {\partial u_2} \\ 
\frac \partial {\partial u_3} \\ 
\frac \partial {\partial u_4}
\end{array}
\right) ,  \label{a5.6}
\end{equation}

The (KS) transformation allows us to write the Laplacian $\overrightarrow{%
\nabla }^2$ in $R^3$ in terms of the laplacian $\Box ^2$ in $R^4$ as

\begin{equation}
\overrightarrow{\nabla }^2=\frac 1{4r}\Box ^2-\frac 1{4r^2}\frac{\partial ^2%
}{\partial x_4^2},\qquad \Box ^2=\sum_{j=1}^4\frac{\partial ^2}{\partial
u_j^2}  \label{a5.7}
\end{equation}
where $r=(x_1^2+x_2^2+x_3^2)^{\frac 12}=u_1^2+u_2^2+u_3^2+u_4^2.$

Using the Schwinger's integral representation, the Green's function relative
to the potential $V_4(\overrightarrow{r})$ is written 
\begin{equation}
G(\overrightarrow{r},\overrightarrow{r^{\prime }};E)\!=\!\int_0^\infty
\!dS\exp \!\left\{ \!-\frac{iS}\hbar \!\left[ \!-\frac{\hbar ^2}{2M}%
\overrightarrow{\nabla }^2\!+\!V_4(\overrightarrow{r})\!-\!E\!-\!i0\!\right]
\!\right\} \delta (\overrightarrow{r}\!-\!\overrightarrow{r^{\prime }}).
\label{a5.8}
\end{equation}
It is possible to introduce an additional variable $x_4$ by means of the
well-known identity 
\begin{equation}
\int_{-\infty }^\infty \exp \left[ \frac i\hbar \left( \frac{\hbar ^2S}{8Mr^2%
}\frac{\partial ^2}{\partial x_4^2}\right) \right] \delta (x_4)dx_4=1,
\label{a5.9}
\end{equation}
and to show that expression (\ref{a5.9}) can be written as 
\begin{equation}
G(\overrightarrow{r},\overrightarrow{r^{\prime }};E)=\int_{-\infty }^\infty
dx_4G(\overrightarrow{r},x_4,\overrightarrow{r^{\prime }},0;E),
\label{a5.10}
\end{equation}
where 
\begin{equation}
G(\overrightarrow{r},x_4,\overrightarrow{r^{\prime }},0;E)=\int_0^\infty
dS\exp \left\{ -\frac{iS}\hbar \left[ H_T-E-i0\right] \right\} \delta (%
\overrightarrow{r}-\overrightarrow{r^{\prime }})\delta (x_4),  \label{a5.11}
\end{equation}
with 
\begin{equation}
H_T=-\frac{\hbar ^2}{2M}\left( \overrightarrow{\nabla }^2+\frac 1{2r^2}\frac{%
\partial ^2}{\partial x_4^2}\right) +V_4(\overrightarrow{r}).  \label{a5.12}
\end{equation}

Using (\ref{a5.7}), (\ref{a5.3}), the Jacobian of this transformation given
by \newline
$\frac{\partial (x_1,x_{2,}x_{3,}x_4)}{\partial (u_1,u_2,u_3,u_4)}=16r^2 $,
and the time transformation $\tau =\frac S{4r}$, the Green's function (\ref
{a5.11} ) can be put in the form 
\begin{eqnarray}
G(\overrightarrow{r},x_4,\overrightarrow{r^{\prime }},0;E) &=&\frac
1{4r}\int_0^\infty d\tau \exp \left[ \frac i\hbar (4\alpha _0+i0)\tau \right]
\nonumber \\
&&\times \exp \left\{ -\frac{i\tau }\hbar \left[ -\frac{\hbar ^2}{2M}\Box
^2+V(u)\right] \right\} \prod_{j=1}^4\delta (u_j-u_j^{\prime }),  \nonumber
\\
&&  \label{a5.13}
\end{eqnarray}
where 
\begin{eqnarray}
V(u) &=&\frac{\hbar ^2k_1^2}{2M}\left( \frac 1{u_1^2+u_2^2}-\frac
1{u_3^2+u_4^2}\right) +\frac{\hbar ^2}{2M}\left( \frac 1{u_1^2+u_2^2}\right.
\nonumber \\
&&\left. +\frac 1{u_3^2+u_4^2}\right) F\left( \frac{u_2u_3-u_1u_4}{%
u_1u_3+u_2u_4}\right) -4Er.  \label{a5.14}
\end{eqnarray}
The evaluation of this expression is possible in two coordinate systems.

\subsubsection{Parabolic coordinates}

Going on to the double polar coordinates 
\begin{equation}
\left\{ 
\begin{array}{c}
(u_1,u_2)\rightarrow (\eta ,\phi _1):u_1=\frac \eta {\sqrt{2}}\cos \phi
_1,\quad u_2=\frac \eta {\sqrt{2}}\sin \phi _1, \\ 
(u_3,u_4)\rightarrow (\xi ,\phi _2):u_3=\frac \xi {\sqrt{2}}\cos \phi
_2,\quad u_4=\frac \xi {\sqrt{2}}\sin \phi _2,
\end{array}
\right.  \label{a5.15}
\end{equation}
the Green's function ( \ref{a5.13} ) becomes 
\begin{eqnarray}
G(\overrightarrow{r},x_4,\overrightarrow{r^{\prime }},0;E) &=&\frac
1{2r}\int_0^\infty d\tau \exp \left[ \frac i\hbar (2\alpha _0+i0)\tau \right]
\exp \left[ -\frac{i\tau }\hbar H_p\right]  \nonumber \\
&&\times \frac{\delta (\xi -\xi ^{\prime })\delta (\eta -\eta ^{\prime })}{%
\sqrt{\xi \xi ^{\prime }\eta \eta ^{\prime }}}\delta (\phi _1-\phi
_1^{\prime })\delta (\phi _2-\phi _2^{\prime }),  \label{a5.16}
\end{eqnarray}
where 
\begin{equation}
H_p=-\frac{\hbar ^2}{2M}\left( \frac{\partial ^2}{\partial \xi ^2}+\frac{%
\partial ^2}{\partial \eta ^2}+\frac 1\xi \frac \partial {\partial \xi
}+\frac 1\eta \frac \partial {\partial \eta }+\frac 1{\xi ^2}\frac{\partial
^2}{\partial \phi _1^2}+\frac 1{\eta ^2}\frac{\partial ^2}{\partial \phi _2^2%
}\right) +V_p,  \label{a5.17}
\end{equation}
with 
\begin{equation}
V_p=-\frac{\hbar ^2k_1^2}{2M}\left( \frac 1{\xi ^2}-\frac 1{\eta ^2}\right) +%
\frac{\hbar ^2}{2M}\left( \frac 1{\xi ^2}+\frac 1{\eta ^2}\right) F(\tan
(\phi _1-\phi _2)-E(\xi ^2+\eta ^2).  \label{a5.18}
\end{equation}
Here we have applied the rescaling of $2\tau $ to $\tau $.

At this point, we notice that the separation of variables is not possible.
To achieve it, it is necessary to perform the integration on the variable $%
x_4$ by using the Euler's angles $\phi _1=\frac{\alpha +\phi }2,$ $\phi _2=%
\frac{\alpha -\phi }2,$ ($0\leq \phi <2\pi ,\quad 0\leq \alpha <4\pi $)$.$

Then, it is easy to see that $dx_4=rd\alpha $ and that by integration over $%
\alpha $ we are led to $4\pi \delta _{\nu ,0}$. It follows that the Green's
function (\ref{a5.9}) can be written 
\begin{equation}
G(\overrightarrow{r},\overrightarrow{r^{\prime }};E)=\int_0^\infty d\tau
\exp \left[ \frac i\hbar (2\alpha _0+i0)\tau \right] K_{\xi \eta }(\phi
,\phi ^{\prime };\tau )K(\xi ,\eta ,\xi ^{\prime },\eta ^{\prime };\tau ),
\label{a5.19}
\end{equation}
where 
\begin{equation}
K_{\xi \eta }(\phi ,\phi ^{\prime };\tau )=\exp \left\{ -\frac{i\tau }\hbar 
\frac{\hbar ^2}{2M}\left( \frac 1{\xi ^2}+\frac 1{\eta ^2}\right) \left[ -%
\frac{\partial ^2}{\partial \phi ^2}+F(\tan \phi )\right] \right\} \delta
(\phi -\phi ^{\prime }),  \label{a5.20}
\end{equation}
and 
\begin{equation}
K(\xi ,\eta ,\xi ^{\prime },\eta ^{\prime };\tau )=\frac 1{\sqrt{\xi \xi
^{\prime }\eta \eta ^{\prime }}}\exp \left[ -\frac{i\tau }\hbar H(\xi ,\eta )%
\right] \delta (\xi -\xi ^{\prime })\delta (\eta -\eta ^{\prime }),
\label{a5.21}
\end{equation}
with 
\begin{equation}
H(\xi ,\eta )=-\frac{\hbar ^2}{2M}\left( \frac{\partial ^2}{\partial \xi ^2}+%
\frac{\partial ^2}{\partial \eta ^2}\right) -\frac{\hbar ^2}{2M}\left(
k_1^2-\frac 14\right) \left( \frac 1{\xi ^2}-\frac 1{\eta ^2}\right) -4E(\xi
^2+\eta ^2).  \label{a5.22}
\end{equation}

In order to bring to a constant the mass appearing in the Hamiltonian
contained the kernel to expression (\ref{a5.20}) let's perform the time
transformation $\tau \rightarrow \sigma $ defined by $\sigma =\tau \left(
\frac 1{\xi ^2}+\frac 1{\eta ^2}\right) .$ Then, it follows that 
\begin{eqnarray}
K_{\xi \eta }(\phi ,\phi ^{\prime };\tau ) &=&\exp \left\{ -\frac{i\sigma }%
\hbar \frac{\hbar ^2}{2M}\left[ -\frac{\partial ^2}{\partial \phi ^2}+F(\tan
\phi )\right] \right\} \delta (\phi -\phi ^{\prime })  \nonumber \\
&=&\int dE_{\lambda _\phi }\exp \left[ -\frac i\hbar \left( \frac 1{\xi
^2}+\frac 1{\eta ^2}\right) E_{\lambda _\phi }\tau \right] \Psi _{\lambda
_\phi }(\phi )\Psi _{\lambda _\phi }^{*}(\phi ^{\prime }),  \nonumber \\
&&  \label{a5.23}
\end{eqnarray}
with $E_{\lambda _\phi }=\frac{\hbar ^2\lambda _\phi ^2}{2M}.$ Here, we have
assumed that the propagator associated with the potential $F(\tan \phi )$ is
known.

Let's now insert (\ref{a5.23}) and (\ref{a5.21}) in the expression (\ref
{a5.19}), we obtain 
\begin{eqnarray}
G(\overrightarrow{r},\overrightarrow{r^{\prime }};E) &=&\int dE_{\lambda
_\phi }\Psi _{\lambda _\phi }(\phi )\Psi _{\lambda _\phi }^{*}(\phi ^{\prime
})\frac 1{\sqrt{\xi \eta \xi ^{\prime }\eta ^{\prime }}}\int_0^\infty d\tau
\exp \left[ \frac i\hbar (2\alpha _0+i0)\tau \right]  \nonumber \\
&&\times K(\xi ,\xi ^{\prime };\tau )K(\eta ,\eta ^{\prime };\tau ),
\label{a5.24}
\end{eqnarray}
where 
\begin{eqnarray}
K(u,u^{\prime };\tau ) &=&\exp \left\{ -\frac{i\tau }\hbar \left[ -\frac{%
\hbar ^2}{2M}\left( \frac{\partial ^2}{\partial u^2}-\frac{\lambda _\phi
^2\mp k_1^2-\frac 14}{u^2}\right) -Eu^2\right] \right\} \delta (u-u^{\prime
})  \nonumber \\
&=&\exp \left\{ -\frac{i\tau }\hbar \left[ T_1(u)+2\hbar ^2\omega ^2T_3(u)%
\right] \right\} \delta (u-u^{\prime });(u\equiv \xi ,\eta ),  \label{a5.25}
\end{eqnarray}
with $\omega =\sqrt{-\frac{2E}M}.$

Following formula (\ref{a2.13}), the kernels (\ref{a5.25}) are written 
\begin{equation}
K(u,u^{\prime };\tau )=\frac{M\omega }{i\hbar \sin (\omega \tau )}\exp \left[
\frac{iM\omega }{2\hbar }(u^2+u^{\prime 2})\cot (\omega \tau )\right]
I_{\lambda _{\mp }}\left( \frac{M\omega uu^{\prime }}{i\hbar \sin (\omega
\tau )}\right) ,  \label{a5.26}
\end{equation}
with $\lambda _{\mp }=\sqrt{\lambda _\phi ^2\mp k_1^2}.$

Inserting (\ref{a5.26}) into (\ref{a5.24}) and by a procedure similar to
that which has led to result (\ref{a4.16}), we obtain 
\begin{eqnarray}
G(\overrightarrow{r},\overrightarrow{r^{\prime }};E) &=&i\hbar \int
dE_{\lambda _\phi }\left\{ \sum_{n_1,n_2=0}^\infty \frac{\Psi
_{n_1,n_2,\lambda _\phi }(\xi ,\eta ,\phi )\Psi _{n_1,n_2,\lambda _\phi
}^{*}(\xi ^{\prime },\eta ^{\prime },\phi ^{\prime })}{E+i0-E_N}\right. 
\nonumber \\
&&\left. +\int_0^\infty dp\int_{-\infty }^\infty d\varsigma \frac{\Psi
_{p,\varsigma ,\lambda _\phi }(\xi ,\eta ,\phi )\Psi _{p,\varsigma ,\lambda
_\phi }^{*}(\xi ^{\prime },\eta ^{\prime },\phi ^{\prime })}{E+i0-E_p}%
\right\} ,  \label{a5.27}
\end{eqnarray}
with the normalized wave functions and the energy spectrum for

-the bound states: 
\begin{eqnarray}
\Psi _{n_1,n_2,\lambda _\phi }(\xi ,\eta ,\phi ) &=&\left[ \frac 2{a^3N^4}%
\frac{n_1!n_2!}{\Gamma (n_1+\lambda _{-}+1)\Gamma (n_2+\lambda _{+}+1)}%
\right] ^{\frac 12}\left( \frac \xi {aN}\right) ^{\frac{\lambda _{-}}2} 
\nonumber \\
&&\times \left( \frac \eta {aN}\right) ^{\frac{\lambda _{+}}2}\exp \left( -%
\frac{\xi ^2+\eta ^2}{2aN}\right) L_{n_1}^{\lambda _{-}}\left( \frac{\xi ^2}{%
aN}\right) L_{n_2}^{\lambda _{+}}\left( \frac{\eta ^2}{aN}\right)  \nonumber
\\
&&\times \Psi _{\lambda _\phi }(\phi ),  \label{a5.28}
\end{eqnarray}

\begin{equation}
E_N=-\frac{M\alpha _0^2}{2\hbar ^2N^2},\quad N=n_1+n_2+\frac 12(\lambda
_{-}+\lambda _{+})+1,  \label{a5.29}
\end{equation}

-the continuous states: 
\begin{eqnarray}
\Psi _{p,\varsigma ,\lambda _\phi }(\xi ,\eta ,\phi ) &=&\frac{\left| \Gamma
\left( \frac 12(1+\lambda _{+})+\frac i{2p}(\frac 1a+\varsigma )\right)
\Gamma \left( \frac 12(1+\lambda _{-})+\frac i{2p}(\frac 1a-\varsigma
)\right) \right| }{2\pi \Gamma (1+\lambda _{+})\Gamma (1+\lambda _{-})} 
\nonumber \\
&&\times \frac{e^{\frac \pi {2ap}}}{\sqrt{p}\xi \eta }M_{-\frac i{2p}(\frac
1a+\varsigma ),\frac{\lambda _{-}}2}(-ip\xi ^2)M_{-\frac i{2p}(\frac
1a-\varsigma ),\frac{\lambda _{+}}2}(-ip\eta ^2)  \nonumber \\
&&\times \Psi _{\lambda _\phi }(\phi ).  \label{a5.30}
\end{eqnarray}
\begin{equation}
E_p=\frac{\hbar ^2p^2}{2M}.  \label{a5.31}
\end{equation}

\subsubsection{Spherical coordinates}

With the help of the change of variables defined by 
\begin{equation}
(\xi ,\eta )\rightarrow (\sqrt{2r}\cos \frac \theta 2,\sqrt{2r}\sin \frac
\theta 2),  \label{a5.32}
\end{equation}
and by applying the Bateman 's expansion formula (\ref{a3.13}), expression (%
\ref{a5.24}) is rewritten 
\begin{eqnarray}
G(\overrightarrow{r},\overrightarrow{r^{\prime }};E) &=&\int dE_{\lambda
_\phi }\Psi _{\lambda _\phi }(\phi )\Psi _{\lambda _\phi }^{*}(\phi ^{\prime
})\sum_{n=0}^\infty \Phi _n^{(\lambda _{+},\lambda _{-})}\left( \frac \theta
2\right)  \nonumber \\
&&\times \Phi _n^{(\lambda _{+},\lambda _{-})}\left( \frac{\theta ^{\prime }}%
2\right) \frac 1{\sqrt{\sin \theta \sin \theta ^{\prime }}}\int_0^\infty
d\tau \frac{M\omega }{i\hbar \sin (\omega \tau )}  \nonumber \\
&&\times \exp \left[ \frac i\hbar (2\alpha _0+i0)\tau \right] \exp \left[ 
\frac{iM\omega }\hbar (r+r^{\prime })\cot (\omega \tau )\right]  \nonumber \\
&&\times I_{2n+\lambda _{+}+\lambda _{-}+1}\left( \frac{2M\omega \sqrt{%
rr^{\prime }}}{i\hbar \sin (\omega \tau )}\right) .  \label{a5.33}
\end{eqnarray}

Performing the integration over the time variable $\tau $ with the help of
formula (\ref{a3.17}), we obtain the following final form 
\begin{eqnarray}
G(\overrightarrow{r},\overrightarrow{r^{\prime }};E) &=&\int dE_{\lambda
_\phi }\Psi _{\lambda _\phi }(\phi )\Psi _{\lambda _\phi }^{*}(\phi ^{\prime
})\sum_{n=0}^\infty \frac 1{\sqrt{\sin \theta \sin \theta ^{\prime }}}\Phi
_n^{(\lambda _{+},\lambda _{-})}\left( \frac \theta 2\right)  \nonumber \\
&&\times \Phi _n^{(\lambda _{+},\lambda _{-})}\left( \frac{\theta ^{\prime }}%
2\right) \frac 1{2i\omega rr^{\prime }}\frac{\Gamma (p+n+\frac 12(\lambda
_{+}+\lambda _{-})+1)}{\Gamma (2n+\lambda _{+}+\lambda _{-}+2)}  \nonumber \\
&&\times M_{-p,n+\frac 12(\lambda _{+}+\lambda _{-}+1)}\left( \frac{2M\omega 
}\hbar r^{\prime }\right)  \nonumber \\
&&\times W_{-p,n+\frac 12(\lambda _{+}+\lambda _{-}+1)}\left( \frac{2M\omega 
}\hbar r\right) ,  \label{a5.34}
\end{eqnarray}
where $r>r^{\prime }$ and $p=-\frac{\alpha _0}{\hbar \omega }.$

Following the calculation procedure in subsection (IV.2), one shows that 
\begin{eqnarray}
G(\overrightarrow{r},\overrightarrow{r^{\prime }};E) &=&i\hbar \int
dE_{\lambda _\phi }\sum_{n=0}^\infty \left\{ \sum_{l=0}^\infty \frac{\Psi
_{l,n,\lambda _\phi }(r,\theta ,\phi )\Psi _{l,n,\lambda _\phi
}^{*}(r^{\prime },\theta ^{\prime },\phi ^{\prime })}{E+i0-E_N}\right. 
\nonumber  \label{a5.34} \\
&&\ \left. +\int_0^\infty dk\frac{\Psi _{k,n,\lambda _\phi }(r,\theta ,\phi
)\Psi _{k,n,\lambda _\phi }^{*}(r^{\prime },\theta ^{\prime },\phi ^{\prime
})}{E+i0-\frac{\hbar ^2k^2}{2M}}\right\} ,  \label{a5.35}
\end{eqnarray}
where the normalized wave functions and the energy spectrum are given by 
\begin{eqnarray}
\Psi _{l,n,\lambda _\phi }(\overrightarrow{r}) &=&\frac 1{a\left( l+\lambda
_1+\frac 12\right) ^2}\left[ \frac{2l!}{a\Gamma (l+2\lambda _1+1)}\right]
^{\frac 12}\left( \frac{2r}{a\left( l+\lambda _1+\frac 12\right) }\right)
^{\lambda _1-\frac 12}  \nonumber \\
&&\ \times \exp \left( -\frac r{a\left( l+\lambda _1+\frac 12\right)
}\right) L_l^{2\lambda _1}\left( \frac{2r}{a\left( l+\lambda _1+\frac
12\right) }\right) \sqrt{\frac 2{\sin \theta }}  \nonumber \\
&&\ \times \Phi _n^{(\lambda _{+},\lambda _{-})}(\frac \theta 2)\Psi
_{\lambda _\phi }(\phi ),  \label{a5.36}
\end{eqnarray}
\begin{equation}
E_N=-\frac{M\alpha _0^2}{2\hbar ^2N^2},\quad N=l+\lambda _1+\frac 12\quad 
\text{ and \quad }\lambda _1=n+\frac 12(\lambda _{+}+\lambda _{-}+1),
\label{a5.37}
\end{equation}
for the bound states and 
\begin{eqnarray}
\Psi _{k,n,\lambda _\phi }(\overrightarrow{r}) &=&\frac{\left| \Gamma \left(
\frac 12+\lambda _1-\frac i{ak}\right) \right| }{\Gamma (2\lambda _1+1)}%
\frac{e^{\frac \pi {2ak}}}rM_{-\frac i{ak},\lambda _1}(-2ikr)\frac 1{\sqrt{%
2\sin \theta }}\Phi _n^{(\lambda _{+},\lambda _{-})}\left( \frac \theta
2\right)  \nonumber \\
&&\ \times \Psi _{\lambda _\phi }(\phi ),  \label{a5.38}
\end{eqnarray}
\begin{equation}
E_k=\frac{\hbar ^2k^2}{2M},  \label{a5.39}
\end{equation}
for the continuous states.

\section{Conclusion}

In this paper, we have analyzed through the Milshtein and Strakhovenko
variant of the so$(2,1)$ Lie algebra a set of potentials belonging to three
different classes of Smorodinsky-Winternitz potentials. The use of the
second order differential operators of this algebra allows to write the
Hamiltonian of these physical systems in form of a linear combination of the
latter. Using the Schwinger's integral representation and with the help of
two Baker-Campbell-Hausdorff formulas allowing the separation of the $T_i$
operators and thus simplifying their action on a Laplace transform of a well
chosen Dirac distribution , we have shown that we can construct the Green's
functions in compact form in different coordinate systems. This method can
be compared to the approach of the Schr\"odinger equation and to the
Feynman's path integral technique. It gives a local view of the problem
under consideration like the Schr\"odinger approach, but its advantage is in
the computation of the explicit and compact form of the Green's function
from which the energy spectrum and the suitably normalized wave functions
are simultaneously extracted for the bound states and for the continuous
states if they exist at one and the same time.

The advantage of the path integral approach in comparison with the algebraic
method is that it provides a global view of the dynamics of the physical
system, but a problem of singularity is often to be found at the origin of
coordinates, and it requires a regularization which is rather complicated to
perform. From this point of view, we can assert that this algebraic method
has the advantage of not presenting this problem owing to the fact that it
is local.

The method of Milshtein and Strakhovenko could become a powerful alternative
approach to the path integral technique if we manage to extend its use to
the treatment of P\"oschl-Teller potential class.

\end{document}